\documentclass[11pt,eprint,nofootinbib]{revtex4-2}
\usepackage[utf8]{inputenc}
\usepackage{geometry, bbold, mathrsfs, tikz, amssymb, simpler-wick, slashed, graphicx, dsfont,float}
\usepackage[italicdiff]{physics}
\usepackage[colorlinks=true, linkcolor=blue, linktocpage=true]{hyperref}

\usepackage[export]{adjustbox}
\numberwithin{equation}{section}

\usepackage[normalem]{ulem}

\definecolor{NRcolor}{RGB}{0, 102, 204}

\newcommand{\nn}[0]{\nonumber \\}

\newcommand{\qwhere}[0]{\quad \text{where} \quad}
\newcommand{\npar}[0]{\bigskip \par \noindent}
\newcommand{\Jfactor}[0]{\frac{\Gamma\left(2-\frac{d}{2}\right) \qty(\Gamma\qty(\frac{d-2}{2}))^2}{(4 \pi)^{\frac{d}{2}} \Gamma(d-2)}}
\tikzstyle{block} = [draw,rectangle,thick,minimum height=2em,minimum width=2em]
\tikzstyle{sum} = [draw,circle,inner sep=0mm,minimum size=2mm]
\tikzstyle{connector} = [->,thick]
\tikzstyle{line} = [thick]
\tikzstyle{branch} = [circle,inner sep=0pt,minimum size=1mm,fill=black,draw=black]
\tikzstyle{guide} = []
\tikzstyle{snakeline} = [connector, decorate, decoration={pre length=0.2cm,
                         post length=0.2cm, snake, amplitude=.4mm,
                         segment length=2mm},thick, blue, ->]
\tikzset{graviton/.style={decorate, decoration={snake, amplitude=.3mm, segment length=1.3mm, pre length=0mm, post length=0mm}, double}}
\usepackage[compat=1.1.0]{tikz-feynman}
\usepackage[compat=1.1.0]{tikz-feynhand}
\definecolor{grayViolet}{RGB}{128,96,154}

\begin{document}

\title{Higher-Derivative Corrections to Reissner--Nordstr\"om Black Holes from Worldline QFT}
\author{Siddarth Ajith} 
\author{Ravisankar Rajagopal} 
\author{Nur Rifat}
\author{Diana Vaman} 
\author{Kent Yagi}

\affiliation{Department of Physics,  University of Virginia, \\ Charlottesville, Virginia 22904-4714, USA}


\date{\today}

\begin{abstract}
    In this paper we derived the  corrections to the Reissner-Nordstr\"om  black hole when higher-derivative $RF^2$ terms (contractions of  the Riemann tensor with the  Maxwell field strength squared) are added to the Einstein-Maxwell action. Such terms arise naturally  in the context of effective field theories. We used wordline QFT methods to obtain the leading order post-Minkowskian corrections. We verified these results by solving the modified Einstein-Maxwell field equations in closed form, to all orders in Newton's constant $G$. We discussed the first law and computed the entropy of the perturbed black holes. 
When fixing the mass to that of the extremal Reissner-Nordstr\"om black hole, the entropy shift is positive if the weak gravity conjecture holds. In fact, this condition is nothing but the requirement that the spacetime with the extremal Reissner-Nordstr\"om mass possesses an outer horizon in higher-derivative gravity. This condition also  rules out Drummond-Hathrell theory.
  \end{abstract}

\maketitle
\flushbottom
\newpage

  \hypersetup{linkcolor=blue}
   \tableofcontents
   
   \newpage

\section{Introduction}Higher-derivative corrections to Einstein–Maxwell theory arise naturally within effective field theory (EFT) as a consequence of quantum effects, such as integrating out massive charged fields. These corrections modify the charged black hole solutions, including Reissner–Nordstr\"om (RN) \cite{Reissner1916, Nordstrom1918}, and alter their extremality condition, near-horizon structure, and photon propagation. String theory provides an additional source of higher-derivative operators, whose precise form depends on the details of the compactification.
Two well-known examples are the Drummond-Hathrell (DH) \cite{DrummondHathrell1980} effective action and the gauge-invariant vector--tensor theory constructed by Horndeski \cite{Horndeski1976}.  Although both fall within the same general class of curvature–field-strength couplings, their origins are distinct: the DH action arises from one-loop  effects, whereas the Horndeski interaction is uniquely selected by the requirement of gauge invariance and second-order field equations.

For concreteness, consider the parametrization of the Einstein--Maxwell EFT
\begin{align}
\mathcal S
=
\int d^4x \sqrt{-g}
\left[
\frac{1}{2\kappa^2}R
-\frac{1}{4}F_{\mu\nu}F^{\mu\nu}
+\mathcal{L}_{\mathrm{hd}}
\right],
\end{align}
where $\kappa = \sqrt{8\pi G}$, $g$ is the metric determinant, $R$ is the Ricci scalar and $F_{\mu\nu}$ is the Maxwell field strength (aka Faraday tensor).
We first write the four-derivative correction Lagrangian as in \cite{KatsMotlPadi2007}\footnote{ The authors of \cite{KatsMotlPadi2007} left out the terms with coefficients $c'_7$ and $c_{10} $.}:
\begin{align}
\mathcal{L}_{\mathrm{hd}}
&=
c_1 R^2
+c_2 R_{\mu\nu}R^{\mu\nu}
+c_3 R_{\mu\nu\rho\sigma}R^{\mu\nu\rho\sigma}\nonumber\\&
+c_4 R F_{\mu\nu}F^{\mu\nu}
+c_5 R_{\mu\nu}F^{\mu}{}_{\rho}F^{\nu\rho}
+c_6 R_{\mu\nu\rho\sigma}F^{\mu\nu}F^{\rho\sigma}\nonumber\\&
+c_7 (F_{\mu\nu}F^{\mu\nu})^2+c'_7 F_{\mu\nu}F^{\nu\rho}F_{\rho\sigma}F^{\sigma\mu}\nonumber\\&
+c_8 (\nabla_\mu F_{\rho\sigma})(\nabla^\mu F^{\rho\sigma})+c_{9} (\nabla_\mu F_{\rho\sigma})(\nabla^\rho F^{\mu\sigma})+c_{10} \nabla_\rho F^{\rho\mu} \nabla_\sigma F^{\sigma}{}_{\mu}
,\label{genhd}
\end{align}
where  $c_i$ are coupling constants.

The Drummond-Hathrell action \cite{DrummondHathrell1980}, obtained by 
integrating out a massive charged spin $1/2$ fermion of mass $M$ and charge $Q$ in a curved spacetime, yields the following $RF^2$ coefficients  at one loop: 

\begin{align}
\mathcal{L}_{\mathrm{DH}}
=
-\frac{\hbar Q^2}{4\pi^2 M^2}
\left(
-\frac{1}{144}R\,F_{\mu\nu}F^{\mu\nu}
+\frac{13}{360}R_{\mu\nu}F^{\mu}{}_{\rho}F^{\nu\rho}
-\frac{1}{360}R_{\mu\nu\rho\sigma}F^{\mu\nu}F^{\rho\sigma}
-\frac{1}{30} \nabla_\rho F^{\rho\mu}\nabla_\sigma F^\sigma{}_\mu\right).\label{dh}
\end{align}
In contrast, the Horndeski  interaction \cite{Horndeski1976}, which was derived from the requirement of no ghosts (hence second order equations of motion) and gauge invariance, has the form:
\begin{align}
\mathcal{L}_{\mathrm{Horndeski}}
=
\frac{\gamma}{2}
\left(
R\,F_{\mu\nu}F^{\mu\nu}
-4R_{\mu\nu}F^{\mu}{}_{\rho}F^{\nu\rho}
+R_{\mu\nu\rho\sigma}F^{\mu\nu}F^{\rho\sigma}
\right),\label{horndeski}
\end{align}
with some coupling constant $\gamma$.
Terms of $R^2$ type  arise, for example, when integrating out a massive particle coupled to gravity. The Gauss-Bonnet Lagrangian is defined by selecting $c_2=-4c_1, c_3=c_1$.  Lastly, the $(F^2)^2$ and $F^4$ terms describe light-by-light scattering. Such non-linearities are naturally introduced when  integrating out a charged fermion in Quantum Elelctrodynamics (QED), in which case these terms are referred to as the Euler-Heisenberg Lagrangian \cite{Heisenberg:1936nmg}. The Euler-Heisenberg coefficients at one loop are $c_7=-\frac 1{36}( \hbar/M^4) (Q^2/(4\pi)^2) $, $c_{7'}=-\frac {14}5 c_7$.

We note that the coefficients listed in \eqref{genhd} are basis dependent. The  specific way of writing the terms $(\nabla_. F_{..})\nabla^. F^{..}, \; F_{..} \Box F^{..}$, with various index contractions, can change the coefficients of the $RF^2$ terms after integration by parts (see, for example, \cite{Bastianelli2009}). Therefore, it is important to include the terms quadratic in derivatives of the field strength (i.e.  specify the higher dimensional operator basis) in order to give an unambiguous definition to these coefficients.  
In  particular, the $c_8$-terms in \eqref{genhd} are redundant, since they can be rewritten as $c_{9}$-terms using Bianchi identites ($\nabla_{[\mu} F_{\rho\sigma]}=0$):
\begin{align}
\nabla _{\mu} F_{\rho\sigma} \nabla^\mu F^{\rho\sigma}=-2\nabla_\rho F_{\sigma\mu}\nabla^\mu F^{\rho\sigma}.
\end{align}

Lastly,  we can rewrite the $c_{9}$-terms as $c_{10}$-terms and $RF^2$-terms as follows.  Begin with integration by parts:
\begin{align}
\nabla_\rho F_{\sigma\mu}\nabla^\mu F^{\rho\sigma}=F_{\sigma}{}^{\mu}\nabla_\rho\nabla_\mu F^{\rho\sigma},
\end{align}
where the equality holds under the integral sign,  followed by an interchange of the covariant derivatives
\begin{align}
F_{\sigma}{}^{\mu}\nabla_\rho\nabla_\mu F^{\rho\sigma}=F_{\sigma}{}^{\mu}(\nabla_\mu\nabla_\rho F^{\rho\sigma} + R_{\mu\rho}F^{\rho\sigma}
+ R_{\rho\mu\tau}{}^\sigma F^{\tau\rho}).\label{one}
\end{align}
Then integrate the first term by parts again in \eqref{one}. The result is a $c_{10}$-term. The second term in \eqref{one} is one of the $RF^2$-terms, namely $c_5$.  
The third term appears to introduce a new higher-derivative $RF^2$ contraction; however, it can be rewritten in terms of the existing  $c_5$- and $c_6$-terms using the Bianchi identity for the curvature tensor.

This brings us to the final form of our chosen basis of higher derivative operators:
\begin{align}
\mathcal{L}_{\mathrm{hd}}
&=
a_1 R^2
+a_2 R_{\mu\nu}R^{\mu\nu}
+a_3 R_{\mu\nu\rho\sigma}R^{\mu\nu\rho\sigma}\nonumber\\&
+b_1 (F_{\mu\nu}F^{\mu\nu})^2+b_2 F_{\mu\nu}F^{\nu\rho}F_{\rho\sigma}F^{\sigma\mu}\nonumber\\&+\lambda_1 R_{\mu\nu\rho\sigma}F^{\mu\nu}F^{\rho\sigma}\nonumber
+\lambda_2 R_{\mu\nu}F^{\mu}{}_{\rho}F^{\nu\rho}
+\lambda_3 R F_{\mu\nu}F^{\mu\nu}\\&+\lambda_4 \,\nabla_\rho F^{\rho\mu} \nabla_\sigma F^{\sigma}{}_{\mu}
,\label{genhd1}
\end{align}
 with coupling constants $a_i$, $b_i$ and $\lambda_i$.
 The terms listed in \eqref{genhd1} are parity-even. In general, there may be additional parity-odd higher-derivative terms.
For the leading order corrections to electric or magnetic charged black holes the parity-odd terms will not contribute, and we will not consider them here. 
Since the photon-photon interaction terms, $b_1(F^2)^2$ and $b_2F^4$,  contribute to higher order in perturbation theory (order $Q^4$ in the charge) we do not consider them either. 
And, as we will see, the $\lambda_4$ term does not contribute to the leading order charged black hole corrections (which is why it was ignored in  \cite{KatsMotlPadi2007}, though they do keep the redundant $c_8 $ and $c_9$ terms). To order $Q^2$, the relevant terms in \eqref{genhd1} are the $a_i$ and $\lambda_{1,2,3}$ terms.

Previously, corrections due to the four-derivative terms in \eqref{genhd}  to static, charged black hole solutions were considered in \cite{KatsMotlPadi2007}. In particular, the authors explicitly computed only the metric component $g^{rr}$. The effect of the $R^2$ terms on charged black hole solutions  was  explored in an asymptotic flat space in \cite{Campanelli:1994jt}, and in a holographic context in 
\cite{Cremonini:2009ih}. 
On the other hand, working within Einstein-Maxwell theory and using QFT methods, the RN solution was derived perturbatively up to third post-Minkowskian (PM) order by computing the classical contribution of the off-shell currents describing the emission of either a graviton or a photon from a massive charged scalar field, up to two loops \cite{DOnofrio:2022cvn}. Further quantum corrections  (order $\hbar$)  to the RN geometry were obtained in \cite{DonoghueGarbrecht2002}, also in the framework of Einstein-Maxwell theory. The 2PM classical Hamiltonian of a two-body system of charged compact masses was derived recently with a combination of scattering amplitudes and effective field theory methods in \cite{Alonzo-Artiles:2026wbe}. 

Here we are interested in evaluating the effect of the $RF^2$ terms in Eq.~\eqref{genhd1} on the RN black hole. To this end, we employed two methods to find the black hole solutions: the classical limit of scattering amplitudes \cite{Mougiakakos_2021, DOnofrio:2022cvn} as computed in the worldline formalism \cite{Mogull:2020sak} (in Sections~\ref{sec:worldline_review}, \ref{sec:worldline_diagrams} and~\ref{sec:BH_from_worldline}), and directly solving the coupled Einstein-Maxwell equations,  in the presence of the higher-derivative $RF^2$ terms (in Section~\ref{sec:GR_method}).  

In Section~\ref{sec:thermodynamics} we discuss the thermodynamic properties of the perturbed black holes.  We find that the first law holds, though the entropy is no longer given by the Bekenstein-Hawking formula. Instead, it receives corrections which are proportional to  a linear combination of the $\lambda_i$ coefficients and the charge squared, and inversely proportional to the area of the horizon. The same result is obtained in Appendix \ref{AppendixIW}, using the Iyer-Wald formula.
We then compute the entropy and horizon temperature of a black hole that has mass equal to the mass of the extremal RN black hole. We find that positivity and {\it reality} of the entropy shift and  temperature are equivalent to the weak gravity conjecture. We also notice that the weak gravity conjecture is nothing but the demand that the solution with mass equal to that of the extremal RN mass does not have a naked singularity. We also compute the entropy of the extremal black hole in the full theory and find that the change in entropy relative to the extremal RN entropy is independent from the condition imposed by the weak gravity conjecture, and the sign is undetermined.


Lastly, in Section~\ref{sec:BH_shadow}, we use black hole shadow observations of Sgr A*~\cite{EventHorizonTelescope:2022wkp} to constrain the $RF^2$ theory. In particular, we present bounds on the DH and Horndeski combinations of coupling constants as a function of the black hole charge.

\section{Review of the Worldline Formalism }
\label{sec:worldline_review}

We begin with a quick review of the worldline diagrammatic rules which are relevant for this article \cite{Mogull:2020sak, Ajith:2024fna,Du:2024rkf}. Wordline formalism is the first quantized version of quantum field theory \cite{Strassler:1992zr, Schubert_2001, edwards2019quantummechanicalpathintegrals}. Here, one quantizes the trajectory (or the worldline) of the particle, in the presence of a background field, like electromagnetism or gravity. The amplitude for a scalar particle interacting  with $N$ background field excitations can be written as 
\begin{equation}\label{eq-wlamplitude}
\begin{aligned}
    \mathcal{M}_N\left(p, p^{\prime}\right)= \lambda^N & \lim _{\substack{\tau_{N+1} \rightarrow \infty \\
    \tau_0 \rightarrow-\infty}} e^{-i\left(\bar{M}^2-i \epsilon\right)\left(\tau_{N+1}-\tau_0\right)}\left(\prod_{j=1}^{N-1} \int_{-\infty}^{\infty} d \tau_j e^{-\epsilon\left|\tau_j\right|}\right) \\
    & \times\left\langle\mathcal{T}\left(\hat{V}_{\text {out }}\left(\bar{p}^{\prime}, \tau_{N+1}\right) \hat{V}_N\left(\bar{k}_1, \tau_1\right) \hat{V}_2\left(\bar{k}_2, \tau_2\right) \cdots \hat{V}_1\left(\bar{k}_N, 0\right) \hat{V}_{\text {in }}\left(\bar{p}, \tau_0\right)\right)\right\rangle,
\end{aligned}
\end{equation}
where the $\mathcal T(\dots)$ denotes time-ordering,  $\lambda^N$ is a placeholder for the coupling constants with the $N$ background quanta and $V$'s are called the \emph{vertex operators}. The interactions with the background can happen at arbitrary times, which is why we have time ordering, as well as a sum over all possible interaction times. Using the translation invariance of the worldline we set  $\tau_N = 0$.  We defined ``barred" quantities with dimensions of inverse length (we are taking $c=1$, so $\hbar$ has units of mass times length):  
\begin{equation}
    \bar M\equiv \frac{M}\hbar, \qquad \bar p \equiv\frac{p}\hbar,\qquad \bar k\equiv\frac{k}\hbar.
\end{equation}
where $p$ and $p'$ are, respectively, the incoming and outgoing momenta of the particle, and $M$ is the particle's mass. The momenta of the background gravitons or photons  which are absorbed/emitted from the worldline are denoted by $k$. 
The dimension of the worldline ``time" $\tau$ is length squared.
The dimensionless coupling for the emission of a photon from the  worldline is
\begin{equation}
    \lambda\equiv\bar Q=\frac {Q}{\sqrt \hbar},
\end{equation}
where $Q$ is a multiple of the electron charge $e$, and for the emission of a graviton, the coupling constant is 
\begin{equation}
    \lambda\equiv\bar \kappa=\sqrt{8\pi\hbar G},
\end{equation}
where $G$ is the four dimensional  Newton's constant, with  units of length over mass.  $\bar \kappa$ has dimensions of length\footnote{We recall that the Einstein-Hilbert action is conventionally normalized as $S_{\text{EH}}=\frac{1}{2\kappa^2}\int d^4 x \sqrt{-g} R$ and has units of $\hbar$. Here we find it convenient to absorb $\hbar$ into the coupling constants and work with QFT actions which are dimensionless, so the partition function is e.g. $\exp(i S_{\text {EH}})$ rather than $\exp((i/\hbar) S_{\text {EH}})$. Therefore our normalization is  $S_{\text {EH}}=\frac{1}{2\bar\kappa^2}\int d^4 x \sqrt{-g} R$. As a consequence, all the graviton $\bar h_{\mu\nu}$ vertices, with $\bar \kappa \bar h_{\mu\nu}\equiv g_{\mu\nu}-\eta_{\mu\nu}$, come with factors of $\bar\kappa$. E.g. a graviton $n$-point vertex has a $\bar\kappa^{n-2}$-factor dependence.}.  

The in/out on-shell scalar vertex operators are given by
\begin{equation}
    \hat{V}_{\text {in }}\left(\tau_0\right)=\exp \left(i \bar{p} \cdot \hat{x}\left(\tau_0\right)\right), \quad p^2=- M^2, \quad \hat{V}_{\text {out }}\left(\tau_{N+1}\right)=\exp \left(-i \bar{p}^{\prime} \cdot \hat{x}\left(\tau_{N+1}\right)\right), \quad p^{\prime 2}=-M^2.
\end{equation}
They create on-shell scalar particles at asymptotic times. The vertices $V_i$ denote the emission/absorption of background field quanta. Including the corresponding couplings into the definition of the vertex operators, we have 
\begin{equation}
    \hat{V_i}(\bar{k}, \tau)=\left(\begin{array}{c}
                -i \bar{Q} \varepsilon_\mu \dot{\hat{x}}^\mu(\tau_i) \\
                -\frac{i}{2} \bar{\kappa} \varepsilon_{\mu \nu}\dot{\hat{x}}^\mu(\tau_i)\dot{\hat{x}}^\nu(\tau_i)
                \end{array}\right) e^{i \bar{k} \cdot\hat{x}(\tau_i)},
\end{equation}
where the entries represent a 
photon and graviton emission vertex,  respectively. The ``polarizations" $\varepsilon_\mu$ and $\varepsilon_{\mu\nu}$ are placeholders for the photon and graviton vertex index structure. The $\hat x(\tau)$ are coordinates of the quantized worldline, and overdots denote $\tau$-derivatives. The expectation value is computed using Wick's theorem and the (infinite) worldline Green's function, which we henceforth referred to as ``an $\langle x x\rangle$ contraction on the worldline":
\begin{equation}
    \left\langle\mathcal{T} \hat x^\mu\left(\tau_i\right) \hat x^\nu\left(\tau_j\right)\right\rangle=-\frac{i}{2} \eta^{\mu \nu}\left|\tau_i-\tau_j\right|.
\end{equation}
In order to take the classical limit of the worldline, one first performs all the contractions with the in/out vertices and take the limits to obtain
\begin{equation}\label{eq-wl-with-inout-contractions-done}
   {\mathcal{M}}_N\left(p, p^{\prime}\right)=\left(\prod_{j=1}^{N-1} \int_{-\infty}^{\infty} d \tau_j e^{-\epsilon\left|\tau_j\right|}\right)\left\langle\mathcal{T}\left(\hat{V}_1\left(\bar{k}_1, \tau_1\right) \hat{V}_2\left(\bar{k}_2, \tau_2\right) \cdots \hat{V}_N\left(\bar{k}_N, 0\right)\right)\right\rangle
\end{equation}
where the emission vertex operators have changed to
\begin{equation}
    \hat{V}(\bar k, \tau)=\left(\begin{array}{c}
                -i \bar{Q} \varepsilon \cdot(\bar P+\dot{\hat{x}}(\tau)) \\
                -\frac{i}{2} \bar{\kappa} \varepsilon_{\mu \nu}\left(\bar P^\mu+\dot{\hat{x}}^\mu(\tau)\right)\left(\bar P^\nu+\dot{\hat{x}}^\nu(\tau)\right)
            \end{array}\right) e^{i \bar{k} \cdot(\bar P\tau+\hat{x}(\tau))} \qwhere \bar P \equiv \frac{\bar{p} + \bar p'}{2}\,.
\end{equation}
The amplitude in \eqref{eq-wl-with-inout-contractions-done} is represented by the worldline diagram
\begin{equation}
   {\mathcal{M}}_N\left(p, p^{\prime}\right) = \includegraphics[valign=c,scale=1.25]{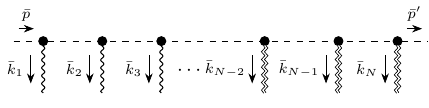},
\end{equation}
where the single-wavy line corresponds to a background photon and the double-wavy line corresponds to a background graviton.
The blobs at vertices are there to remind us that the vertices can slide, due to the integration over the worldline parameter $\tau$. The worldline diagram yields  the {\it exact } same result that one would compute using QFT Feynman rules, just repackaged in a convenient way for the purpose of taking the classical limit.

\npar The classical limit is understood as an $\hbar$ expansion. We work in the limit where the momentum of the source is much larger than the background field momenta. Specifically we take
\begin{equation}
    p^\mu=\hbar \bar p = \text{finite}, \qquad \bar k^\mu = \frac{k^\mu}\hbar=\text{finite.}
\end{equation}
We also rescale the worldline time $\tau \to \tau/\hbar$. This naturally introduces an $\hbar$ in the $\langle xx\rangle$ contractions:
\begin{equation}
     \left\langle\mathcal{T} x^\mu\left(\tau_i\right) x^\nu\left(\tau_j\right)\right\rangle=-\frac{i}{2} \eta^{\mu \nu}\hbar \left|\tau_i-\tau_j\right|.
\end{equation}
and the final form of the worldline amplitude is 
\begin{equation}\label{eq-wl-with-inout-contractions-done-cls}
   {\mathcal{M}}_N\left(p, p^{\prime}\right)=\hbar^{N-1}\left(\prod_{j=1}^{N-1} \int_{-\infty}^{\infty} d \tau_j e^{-\epsilon\left|\tau_j\right|}\right)\left\langle\mathcal{T}\left(\hat{V}_1\left(\bar{k}_1, \tau_1\right) \hat{V}_2\left(\bar{k}_2, \tau_2\right) \cdots \hat{V}_N\left(\bar{k}_N, 0\right)\right)\right\rangle
\end{equation}
with
\begin{equation}
    \hat{V}(\bar k, \tau)=\left(\begin{array}{c}
                -\frac{i}\hbar \bar{Q} \varepsilon \cdot(P+\dot{\hat{x}}(\tau)) \\
                -\frac{i}{2\hbar} \bar{\kappa} \varepsilon_{\mu \nu}\left(P^\mu+\dot{\hat{x}}^\mu(\tau)\right)\left(P^\nu+\dot{\hat{x}}^\nu(\tau)\right)
            \end{array}\right) e^{i \bar{k} \cdot(P \tau+\hat{x}(\tau))} .
\end{equation}
Therefore, the classical limit can be cast into an expansion in the number of contractions between the vertex operators. In the next section, we will elaborate this a bit further, and see which worldline diagrams will actually contribute to the calculation of the metric. 

\section{Classical Spacetimes as Worldline Diagrams}
\label{sec:worldline_diagrams}
\subsection{Einstein's Equations in De Donder Gauge}
Our goal is to find the metric sourced by a particular static energy-momentum source  $T_{\mu \nu}$  by solving the Einstein equations in the presence of higher-derivative corrections:
\begin{equation}
G_{\mu \nu} + ({\text{higher-derivative\;terms}})= {\kappa^2} T_{\mu \nu},
\label{eq:einstein}
\end{equation}
where \begin{equation}
    \kappa = \sqrt{8 \pi G}.
\end{equation}
The units for $T_{\mu\nu}$ are energy density, or, with $c=1$, mass per volume.  We assume that asymptotically the spacetime is flat and we parametrize the metric as
\begin{equation}
g_{\mu \nu} = \eta_{\mu \nu} + \kappa h_{\mu \nu}.
\label{eq:metric_decomposition}
\end{equation}
We gauge fix by imposing the linearized de~Donder gauge condition:
\begin{equation}
\eta^{\mu \nu} \Gamma_{\mu \nu}^\sigma = 0 \Longrightarrow \partial_\nu h^{\nu \sigma} - \frac{1}{2} \partial^\sigma h = 0,
\label{eq:dedonder}
\end{equation}
where $h = h_{\mu \nu} \eta^{\mu \nu}$ is the trace of $h_{\mu \nu}$. In this coordinate system, the Einstein equations can be written as:
\begin{equation}
\square h_{\mu \nu} = -2{\kappa} \left( \mathcal T_{\mu \nu} - \frac{1}{d-1} \eta_{\mu \nu} \mathcal T \right),
\label{eq:box_h}
\end{equation}
where $d$ is the number of {\it{spatial}} dimensions.  $\mathcal T = \mathcal T^{\alpha \beta} \eta_{\alpha \beta}$ is the trace of the energy-momentum tensor, which now includes higher order $\kappa$ terms from expanding the Einstein tensor $G_{\mu \nu}$ as well as the higher-derivative corrections: $\mathcal T_{\alpha\beta}= T_{\alpha\beta}+\mathcal O(\kappa)$.  The idea is to solve the off-shell current $\mathcal T_{\alpha\beta}$ order-by-order in $\kappa$  and in the end evaluate the integrals in \eqref{eq:box_h} to arrive at the metric.

Also, due to the gauge condition \eqref{eq:dedonder}, one can check that $\mathcal T_{\mu\nu}$ is a conserved current
\begin{equation}
    \partial_\mu\mathcal T^{\mu\nu}=0.
\end{equation}

 We will be interested in static solutions to the Einstein equations. Upon Fourier transforming, we obtain:
\begin{equation}
 \kappa h_{\mu \nu}(\boldsymbol{x}) = -2\kappa^2\int \frac{d^d \boldsymbol{q}}{(2\pi)^d} \frac{e^{i \boldsymbol{q} \cdot \boldsymbol{x}}}{\boldsymbol{q}^2}
\left(\mathcal T_{\mu \nu}(\boldsymbol{q}^2) - \frac{1}{d-1} \eta_{\mu \nu} \mathcal T(\boldsymbol{q}^2) \right).
\label{eq:h_fourier}
\end{equation}

It will be useful to parametrize  $\mathcal T_{\mu \nu}$ based on Lorentz covariance and current conservation. 
We begin by expressing $\mathcal T_{\mu\nu}$ in terms of the 
Lorentz covariant momenta $P_\mu$ and $q_\mu$ characterizing the source and graviton, and of course the Minkowski metric.
\label{eq:simplified_tensor}
In the rest frame of the source, where $P_\mu = m \delta_\mu^0$ and with $q^\mu=(0,\boldsymbol{q})$, these conditions lead to: 
\begin{equation}
\mathcal T_{\mu \nu}(\boldsymbol{q}) = C_1(\boldsymbol{q}^2) \delta_\mu^0 \delta_\nu^0 - C_2(\boldsymbol{q}^2) \left( -\eta_{\mu \nu} + \frac{\boldsymbol{q}_\mu \boldsymbol{q}_\nu}{\boldsymbol{q}^2} \right)\,,
\label{eq:form_factors}
\end{equation}
where $C_1$ and $C_2$ are called form factors. In the next section we will determine them for a point-like source of mass $m$ and charge $Q$, to leading order in the higher derivative couplings $\lambda_i$, $i=1,\dots 4$, and evaluate the metric in terms of these form factors:
\begin{equation}
\kappa h_{\mu \nu}(\boldsymbol{x}) = -\frac{\kappa^2}{2} \int \frac{d^d \boldsymbol{q}}{(2\pi)^d} \frac{e^{i \boldsymbol{q} \cdot \boldsymbol{x}}}{\boldsymbol{q}^2} \left( 
C_1(\boldsymbol{q}^2) \left( \delta_\mu^0 \delta_\nu^0 + \frac{\eta_{\mu \nu}}{d-1} \right)
- C_2(\boldsymbol{q}^2) \left( \frac{\boldsymbol{q}_\mu \boldsymbol{q}_\nu}{\boldsymbol{q}^2} + \frac{\eta_{\mu \nu}}{d-1} \right)
\right).
\label{eq:metric_formfactors}
\end{equation}
\label{eq:tmunu_rest_classical}
\label{eq:form_factor_solution}
\subsection{Maxwell's Equations in Lorenz gauge}

Let us now discuss the Maxwell sector.
Similarly to the metric case, in the Lorenz gauge  $\partial_\mu A^\mu=0$, we can write the Maxwell equation as
\begin{equation}
\square A_\mu = \mathcal J_\mu,
\end{equation} 
where the matter current $J_\mu$ is replaced by $\mathcal J_\mu$  due to the non-linearities of the Maxwell's field equations, assuming it is sourced by a charged scalar field in a curved background.

In the static limit, the Maxwell field is obtained from the off-shell current 
\begin{equation}
A_\mu(\boldsymbol{x}) = \int \frac{d^3 \boldsymbol{q}}{(2 \pi)^3} \frac{e^{i \boldsymbol{q} \cdot \boldsymbol{x}}}{\boldsymbol{q}^2} \mathcal J_\mu\left(\boldsymbol{q^2}\right).
\end{equation}
Note that as a consequence of the Lorenz gauge choice, $\mathcal J_\mu$ is a conserved current:
\begin{equation}
    \partial_\mu \mathcal J^\mu=0.
\end{equation}
As we did before, we parametrize $\mathcal J_\mu$ in terms of the Lorentz vectors $P_\mu$ and $q_\mu$. In the rest frame of the source and for static fields, the current conservation yields $\vec{A} = 0$. Therefore, we only need one form factor:
\begin{equation}
A_0(\boldsymbol{x}) = \int \frac{d^3 \boldsymbol{q}}{(2 \pi)^3} \frac{e^{i \boldsymbol{q} \cdot \boldsymbol{x}}}{\boldsymbol{q}^2} \mathcal J_0\left(\boldsymbol{q^2}\right).
\end{equation}

\subsection{Off-Shell Currents  From the Classical Limit of Worldline Diagrams}

Next, let us discuss how we can compute the off-shell current $\mathcal T_{\mu \nu}$ as the classical limit of a quantum field theory (QFT) amplitude.  The linearized energy-momentum tensor $T_{\mu\nu}$ is that of  the black hole which we approximate by a pointlike mass $M$.  This is justified if we are ``far away" from the black hole, where far away means distances much greater than the Compton wavelength of the black hole,
\begin{equation}
r \gg \frac{\hbar}{M}. 
\label{eq:compton_limit}
\end{equation}
Then, we can think of the off-shell current as a three-point amplitude, with two external scalar lines corresponding to the in- and out- states of the black hole, and one off-shell graviton line\footnote{Recall that the linearized GR coupling with $T^{\mu\nu}$ is normalized as $( \kappa/2) \int d^4 x\,T^{\mu\nu} h_{\mu\nu}$. Conventionally the units are $\hbar$, with $T$ of units mass density. Here we use $( \bar \kappa/2) \int d^4 x\,T^{\mu\nu} \bar h_{\mu\nu}$, with $\bar h_{\mu\nu}$ of dimensions inverse length.}. 
The graviton emission vertex is cubic, but,
of course, the gravitons self-interact which modifies the  leading order 3-point function, 
\begin{equation}
i \mathcal{M}(p, p') \bigg|_{\text{leading\; order
}}= -i \frac{\bar{\kappa}}2\langle T_{\mu \nu} \rangle \epsilon^{\mu \nu},
\label{eq:3pt_amplitude0}
\end{equation}where $\epsilon^{\mu \nu}$ is the ``polarization" of the off-shell graviton,
to include multiple graviton emissions from the scalar worldline joined to the off-shell graviton line. This leads to 
\begin{equation}
i \mathcal{M}(p, p') = -i \frac{\bar \kappa} 2 \langle \mathcal T_{\mu \nu} \rangle \epsilon^{\mu \nu}
\label{eq:3pt_amplitude}.
\end{equation}

An identical procedure this time involving the three-point function scalar-scalar-photon, with the photon off-shell, and stripping off the polarization vector $\epsilon_\mu$ will lead to the off-shell current $\mathcal J_\mu$. Since photons and gravitons interact, in general there will be corrections to the leading order current.

Schematically, we compute the following:
\begin{equation}
i \mathcal{M}(p, p') = \includegraphics[valign=c]{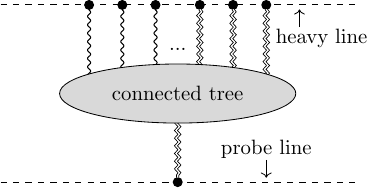}
\label{eq:diagrams}
\end{equation}
where the scalar line with $n$ dangling photons (single wavy-lines) and gravitons (double wavy-lines) is the worldline amplitude computed using the worldline rules as reviewed earlier. This is ``glued" to a ``connected tree" which is a photon-graviton $n$-point function derived from the Einstein-Maxwell (plus possible higher derivative corrections) QFT.  Together with  the connected tree we include the propagators for the photons and gravitons emitted from the worldline(s), and integrate over the loop momenta. In particular we select connected trees which attach to the probe particle worldline with a single graviton or photon propagator. This  propagator is amputated when computing  the  off-shell currents $\mathcal T_{\mu \nu}$ and $\mathcal J_\mu$.

Starting from the dimensionless action $\mathcal S_{\text{EH}}=(1/2\bar\kappa^2)\int d^4x \sqrt{-g} R$ and expanding in fluctuations as
\begin{equation}
g_{\mu\nu}=\eta_{\mu\nu}+\bar\kappa \bar h_{\mu\nu}  
\end{equation} ensures that all bulk graviton $n$-point vertices come with a factor of $\bar\kappa^{n-2}$. Likewise, all ($n$-graviton)-photon-photon vertices will be proportional to  $\bar\kappa^{n}$.

There is one more detail which needs our attention before taking the classical limit. This has to do with the normalization of momentum eigenstates in a relativistic field theory, which have an extra factor of $\sqrt{2E}$ relative to the non-relativistic ones, where $E = \sqrt{\boldsymbol{p}^2 + M^2}$. Accounting for this normalization factor leads to:
\begin{equation}
\frac{1}{\sqrt{4EE'}} i \mathcal{M}(p, p') \equiv -i \frac{\bar\kappa}{2} \langle\mathcal T_{\mu \nu} \rangle \epsilon^{\mu \nu}.
\label{eq:normalized_amplitude}
\end{equation}

 Diagrammatically, the off-shell energy-momentum current is represented as 
\begin{equation}
\langle \mathcal T_{\mu \nu} \rangle = \frac{i}{\tfrac{\bar\kappa}{2} \sqrt{4EE'}} \left(
\includegraphics[valign=c]{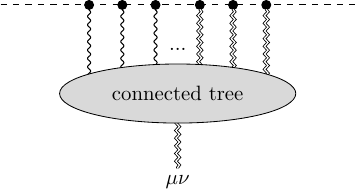}
\label{eq:tmunu_from_diagrams}
\right)
\end{equation}
where  the polarization $\epsilon_{\mu \nu}$ has been removed.

Finally, taking the classical limit of this three-point amplitude (and stripping off the polarization $\epsilon_{\mu \nu}$) will give us the classical energy-momentum tensor:
\begin{equation}
    \mathcal T_{\mu\nu}\equiv\langle \mathcal T_{\mu\nu}\rangle\bigg|_{\text{classical\;limit}}.
\end{equation}
Further substitution in equation~\eqref{eq:box_h} will yield the metric.

We are now ready to extract the  classical ($\hbar\to 0$) part  of the amplitude $\mathcal M$, recalling that we hold fixed $P$, $\bar k_i$, the couplings $\kappa, e$, and $\bar \kappa \bar h_{\mu\nu}$.
The counting is as follows. For an $m$-graviton and $n$-photon worldline, attached to a connected tree with $(m+1)$ gravitons and $n $ photons we have:
\begin{itemize}
    \item $\bar e^n \bar\kappa^m \sim (\sqrt{\hbar})^m \left( \frac{1}{\sqrt{\hbar}} \right)^n \quad$  from the coupling constants of the worldline vertices
    \item $\hbar^{m+n-1}\quad$  from $\tau$ integrals
    \item $\left( \frac{1}{\hbar^2} \right)^m \left( \frac{1}{\hbar} \right)^n \quad$  from the spin dependence of the worldline vertices
    \item $\bar\kappa ^{(m+n+1)-2}\sim (\sqrt{\hbar})^{m+n-1}$  from the connected $(m+n+1)$\;-point graviton/photon tree.
\end{itemize}
Schematically, the classical limit of the off-shell currents is computed without any contractions among the worldline vertices as follows, where the dashed worldline is now dotted, to represent the absence of $\langle xx\rangle $ contractions. For example, the energy-momentum off-shell current is
\begin{equation}
\includegraphics[valign=c]{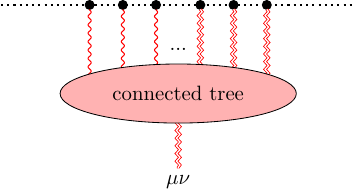}\propto \frac{1}{\hbar \sqrt{\hbar}}
\label{eq:diagram_scaling}
\end{equation}
which is of the desired $\hbar$ order to yield classical contributions to the metric (recall that the metric was expanded in fluctuations as $\bar\kappa\bar h_{\mu\nu}\sim \sqrt{\hbar} \,\bar h_{\mu\nu}$). While radiative correction diagrams (with the background quanta emitted and absorbed by the same scalar line) contribute to the same $\hbar$ order as the connected diagram in  \eqref{eq:diagram_scaling}, it was shown in \cite{Du:2024rkf} that such diagrams lead to scaleless integrals and evaluate to zero in dimensional regularization.
 Each additional $\langle xx\rangle $ contraction among the worldline vertices yields one extra  factor of $\hbar$. Similarly, ``bulk" loops  will introduce additional powers of $\hbar$. Such terms contribute to quantum corrections (see for example \cite{Donoghue1994a} for an earlier evaluation of the quantum corrected Schwarzschild black hole, or \cite{BjerrumBohr2003} for a more recent take on this subject). 

Since our focus is on the classical terms we compute the uncontracted worldline diagrams,  and for simplicity, {\it from now on, we will set $\hbar = 1$. There will be no more distinction made between barred and unbarred quantities in the rest of the paper.}

\bigskip 

The rules for computing the classical limit of worldline diagrams are listed below:
\begin{enumerate}
\item 
For each graviton worldline vertex, include a factor of $-i \kappa P_\alpha P_\beta$, where $P = \frac{p + p'}{2}$.
\item For each photon worldline vertex, include a factor of $-2i Q P_\alpha$.
\item For the $N$ vertices on the worldline (graviton and photon combined), include $N-1$ delta functions, $2\pi \delta(2 P \cdot k_i)$, where $k_i$ is the momentum of the emitted photon or graviton. These follow from the integration over the  emission times $\tau_i$, $i=1,2\dots, N-1$ on the worldline, in the absence of any $\langle xx\rangle$ contractions.
\end{enumerate}
Next we need to glue this to a bulk $N+1$-point connected tree, leaving a dangling graviton or photon line, to obtain the classical off-shell currents
\begin{equation}
 \mathcal T_{\mu \nu}  = \frac{i}{\tfrac{\kappa}{2}\sqrt{4EE'}} \left(\includegraphics[valign=c]{diagrams/Tmunu.pdf}
\label{eq:tmunu_classical}\right)
\end{equation}
and similarly,
\begin{equation}
\mathcal{J}_\mu = \frac{i}{\sqrt{4 E E^{\prime}}} \left(\includegraphics[valign=c]{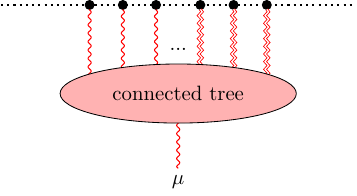}\right)
\end{equation}


\subsection{Warm-up: Order-$G$ Reissner-Nordström Metric  from Worldline QFT}
As a warm-up to the computation of the $RF^2$ effects in \eqref{genhd1} on the black hole geometry, we first use the worldline to compute the leading order terms (in mass and charge) of the Reissner–Nordström black hole. Field theory is well-suited for the perturbative  expansion in the coupling constants, which in the GR literature is known as the Post-Minkowskian (PM) expansion. We evaluate the off-shell currents in a loop expansion (which translates into an expansion in powers of $G$ and $Q^2$). We denote the loop order with a superscript, e.g. $\mathcal T^{(1)}_{\mu\nu}$ denotes a one-loop contribution.

\subsubsection*{Tree Level} 
This is the lowest order in the coupling constant. There is only one diagram that we can draw here,
\begin{equation}
\begin{aligned}
 \mathcal T_{\mu \nu}^{(0)} 
&= \frac{i}{\tfrac{\kappa}{2} (2M)} \left(\includegraphics[valign=c]{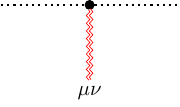}\right) \\
&= \frac{i}{ \kappa M} \left( -i \kappa m^2 \delta_\mu^0 \delta_\nu^0 \right).
\end{aligned}
\label{eq:tree_tmunu}
\end{equation}
Recaling \eqref{eq:form_factors}, we obtained
\begin{equation}
\begin{aligned}
\mathcal T_{00}
(
\boldsymbol{q}^2) &= C_1(\boldsymbol{q}^2) - C_2(\boldsymbol{q}^2) \\
\delta^{ij} \mathcal T_{ij}
(
\boldsymbol{q}^2) &= -C_2(\boldsymbol{q}^2)(1 - d).
\end{aligned}
\label{eq:t_components}
\end{equation}
Comparing with \eqref{eq:tree_tmunu} we identify the leading order form factors:
\begin{equation}
C_1 = M, \quad C_2 = 0.
\label{eq:tree_formfactors}
\end{equation}
Using the integral \eqref{eq:FT} from the appendix, we obtain
\begin{equation}
\kappa h_{\mu \nu}(\boldsymbol{x}) = -\frac{\kappa^2 M}{2} \left( \delta_\mu^0 \delta_\nu^0 + \frac{\eta_{\mu \nu}}{d - 1} \right) \frac{1}{(4\pi)^{d/2}} \Gamma\left( \frac{d - 2}{2} \right) \left( \frac{2}{r} \right)^{d - 2}.
\label{eq:tree_metric}
\end{equation}
Setting  $d=3$ spatial dimensions yields
\begin{equation}
\begin{aligned}
\kappa h_{\mu \nu}(\boldsymbol{x}) &= -\frac{4 G M}{r} \left( 2 \delta_\mu^0 \delta_\nu^0 + \eta_{\mu \nu} \right),
\end{aligned}
\label{eq:schwarzschild_lowest}
\end{equation}
which is the Schwarzschild metric (or Reissner-Nordstr\"om metric) to the lowest order in  $\frac{G M}{r}$.
\subsubsection*{One Loop}
At one loop there are two diagrams:
\begin{equation}
\begin{aligned}
\mathcal T_{\mu \nu}^{(1)} 
&= \frac{i}{ \tfrac{\kappa}{2} 2M} \left(\includegraphics[valign=c]{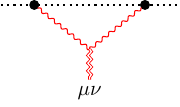}+\includegraphics[valign=c]{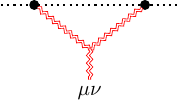}\right) .
\end{aligned}
\end{equation}
We focus on the first diagram to capture the charge  leading-order contribution to the energy-momentum tensor, as the second diagram gives terms of order $\mathcal O((GM)^2)$ and is charge-indepedent:
\begin{equation}
    \begin{aligned}
\mathcal T_{\mu \nu}^{(1)} 
|_{\rm{Q^2 \;term}}
&= \frac{i}{ \kappa M} \frac{1}{2} \int \frac{d^{d+1} k}{(2\pi)^{d+1}} 
\left(-2i Q M \eta^{\alpha 0}\right) \left(-2i Q M \eta^{\beta 0}\right) 
\left( 2\pi \delta(2 M k^0) \right) 
\frac{V_{\mu \nu|\alpha|\beta}(q,k,-(q+k))}{(i k^2)(i(k+q)^2)} \\
&= \frac{i Q^2}{\kappa} \int \frac{d^d \boldsymbol{k}}{(2\pi)^d} 
\frac{V_{\mu \nu|0|0}(\boldsymbol{q},\boldsymbol{k},-(\boldsymbol{q}+\boldsymbol{k}))}{\boldsymbol{k}^2 (\boldsymbol{k}+\boldsymbol{q})^2}\,.
\end{aligned}
\label{eq:one_loop_tmunu}
\end{equation}
The $\frac{1}{2}$ factor is the symmetry factor for the one-loop integral. The Maxwell action,
\begin{equation}
S_{\text{M}} = -\frac{1}{4} \int d^{d+1} x \, \sqrt{g} \, g^{\mu\nu} g^{\rho\sigma} F_{\mu\rho} F_{\nu\sigma}
\label{eq:em_action},
\end{equation}
yields the following cubic photon-photon-graviton vertex
\begin{equation}
V_{\mu \nu|\alpha|\beta}(q,k,k') = -i \frac{\kappa}{2} 2 \Bigg[ \eta^{\rho\sigma} (k_\mu \eta_{\rho \alpha} - k_\rho \eta_{\mu \alpha})(k'_\nu \eta_{\sigma \beta} - k'_\sigma \eta_{\nu \beta}) 
+ \frac{1}{4} \eta_{\mu \nu} (k^\rho \eta_\alpha^\sigma - k^\sigma \eta_\alpha^\rho)(k'_\rho \eta_{\sigma \beta} - k'_\sigma \eta_{\rho \beta}) \Bigg].
\label{eq:vertex}
\end{equation}
From this, we get
\begin{equation}
V_{\mu \nu|0|0}(\boldsymbol{q}, \boldsymbol{k}, -(\boldsymbol{q}+\boldsymbol{k})) 
= -\frac{i \kappa}{2} \left[ 2 \eta_{0 \mu} \eta_{0 \nu} \boldsymbol{k} \cdot (\boldsymbol{k}+\boldsymbol{q}) 
- \left( 2 \boldsymbol{k}_\mu (\boldsymbol{k}+\boldsymbol{q})_\nu 
+ \eta_{\mu \nu} \left( \boldsymbol{k} \cdot (\boldsymbol{k}+\boldsymbol{q}) \right) \right) \right].
\label{eq:vertex_one_photon}
\end{equation}
Substituting in \eqref{eq:one_loop_tmunu} we  find
\begin{equation}
\begin{aligned}
\mathcal T_{00}^{(1)} &= -Q^2 \frac{1}{2} \int \frac{d^d \boldsymbol{k}}{(2\pi)^d} 
\frac{\boldsymbol{k} \cdot (\boldsymbol{k} + \boldsymbol{q})}
{\boldsymbol{k}^2 (\boldsymbol{k} + \boldsymbol{q})^2} \\
\delta^{ij} \mathcal T_{ij}^{(1)} &= Q^2 \frac{d - 2}{2} \int \frac{d^d \boldsymbol{k}}{(2\pi)^d} 
\frac{\boldsymbol{k} \cdot (\boldsymbol{k} + \boldsymbol{q})}
{\boldsymbol{k}^2 (\boldsymbol{k} + \boldsymbol{q})^2}.
\end{aligned}
\label{eq:tmunu_vertex_reduced}
\end{equation}
The integrals are easy to reduce to a master integral. The $\boldsymbol{k}^2$ term in the numerator leads to a scaleless integral, which dimensional regularization sets to zero. Using
\begin{equation}
\boldsymbol{k} \cdot \boldsymbol{q} = \frac{1}{2} \left( (\boldsymbol{k} + \boldsymbol{q})^2 - \boldsymbol{k}^2 - \boldsymbol{q}^2 \right)
\label{eq:kdotq_rewrite}
\end{equation}
and setting scaleless integrals to zero, we arrive at
\begin{equation}
\begin{gathered}
\mathcal T_{00}^{(1)} = \frac{Q^2}{4} \int \frac{d^d \boldsymbol{k}}{(2\pi)^d} 
\frac{\boldsymbol{q}^2}{\boldsymbol{k}^2 (\boldsymbol{k} + \boldsymbol{q})^2} 
= -\frac{Q^2}{4} \vb{q}^2 J_{(1)}(\vb{q}^2)\,,\\
\delta^{ij} \mathcal T_{ij}^{(1)} = -\frac{Q^2 (d - 2)}{4} \int \frac{d^d \boldsymbol{k}}{(2\pi)^d} 
\frac{\boldsymbol{q}^2}{\boldsymbol{k}^2 (\boldsymbol{k} + \boldsymbol{q})^2} 
= -\frac{(d - 2) Q^2}{4}  \vb{q}^2 J_{(1)}(\vb{q}^2)\,.
\end{gathered}
\label{eq:tmunu_integrals_evaluated}
\end{equation}
The integral $ J_{(1)}(\vb{q}^2)$ is a standard one loop integral, and the result is given in the appendix. 
Comparing with \eqref{eq:t_components} yields the form factors,
\begin{equation}
\begin{aligned}
C_1(\boldsymbol{q}^2) &= -\frac{(2d - 3)}{(d - 1)} \frac{Q^2}{4} 
 \vb{q}^2 J_{(1)}(\vb{q}^2)\,, \\
C_2(\boldsymbol{q}^2) &= -\frac{(d - 2)}{(d - 1)} \frac{Q^2}{4} 
 \vb{q}^2 J_{(1)}(\vb{q}^2)\,.
\end{aligned}
\label{eq:formfactors_final}
\end{equation}
\begin{equation}
\begin{aligned}
&\text{We can make use of the integrals in the appendix once more, to evaluate the metric,} \\
&\begin{aligned}
h_{\mu \nu}(\boldsymbol{x}) &= -\frac{\kappa}{2} \int \frac{d^d \boldsymbol{q}}{(2\pi)^d} \frac{e^{i \boldsymbol{q} \cdot \boldsymbol{x}}}{\boldsymbol{q}^2} 
\left( C_1(\boldsymbol{q}^2)\left( \delta_\mu^0 \delta_\nu^0 + \frac{\eta_{\mu \nu}}{d - 1} \right) 
- C_2(\boldsymbol{q}^2) \left( \frac{\boldsymbol{q}_\mu \boldsymbol{q}_\nu}{\boldsymbol{q}^2} -+\frac{\eta_{\mu \nu}}{d - 1} \right) \right)\,, \\
h_{00}(\boldsymbol{x}) &= +\frac{\kappa}{2} \left( \frac{Q^2}{4} \frac{\Gamma\left(2-\frac{d}{2}\right) \qty(\Gamma\qty(\frac{d-2}{2}))^2}{(4 \pi)^{\frac{d}{2}} \Gamma(d-2)} \right) 
\left( \frac{1}{(4\pi)^{d/2}} \frac{\Gamma(d - 2)}{\Gamma\left( \frac{4 - d}{2} \right)} \left( \frac{2}{r} \right)^{2d - 4} \right)
\left( \frac{2(d - 2)}{(d - 1)} \right)\,, \\
h_{ij}(\boldsymbol{x}) &= +\frac{\kappa}{2} \left( \frac{Q^2}{4} \frac{\Gamma\left(2-\frac{d}{2}\right) \qty(\Gamma\qty(\frac{d-2}{2}))^2}{(4 \pi)^{\frac{d}{2}} \Gamma(d-2)} \right) 
\left( \frac{1}{(4\pi)^{d/2}} \frac{\Gamma(d - 2)}{\Gamma\left( \frac{4 - d}{2} \right)} \left( \frac{2}{r} \right)^{2d - 4} \right) \\
&\quad \times \left( \frac{6 - 2d}{(d - 1)(4 - d)} \delta_{ij} 
- \frac{(d - 2)(4 - 2d)}{(d - 1)(4 - d)} \frac{\boldsymbol{x}_i \boldsymbol{x}_j}{r^2} \right)\,,
\end{aligned}
\end{aligned}
\label{eq:metric_final}
\end{equation}
which simplifies to, 
\begin{equation}
\begin{aligned}
\kappa h_{00}(x) &= \frac{\kappa^2 Q^2}{128 \pi^{d - 1}} \frac{2(d - 2)}{(d - 1)} \left( \frac{1}{r} \right)^{2d - 4}\,, \\
\kappa h_{ij}(x) &= \frac{\kappa^2 Q^2}{128 \pi^{d - 1}} 
\left( \frac{(6 - 2d)}{(d - 1)(4 - d)} \delta_{ij} 
+ \frac{2(d - 2)^2}{(d - 1)(4 - d)} \frac{x_i x_j}{r^2} \right) 
\left( \frac{1}{r} \right)^{2d - 4}\,.
\end{aligned}
\label{eq:metric_explicit_final}
\end{equation}
In three spatial dimensions we find,
\begin{equation}
\begin{aligned}
\kappa h_{00}(\boldsymbol{x}) &= \frac{G Q^2}{4 \pi r^2}\,, \\
\kappa h_{ij}(\boldsymbol{x}) &= \frac{G Q^2}{4 \pi r^2} \frac{\boldsymbol{x}_i \boldsymbol{x}_j}{r^2}\,.
\end{aligned}
\label{eq:3d_metric}
\end{equation}

\section{Higher Derivative Couplings}
\label{sec:BH_from_worldline}
In this section, we will add higher derivative couplings to the action of the general type ``$R F^2$'' as in \eqref{genhd1}. The procedure to find the metric at one loop stays almost the same. In the previously derived  \eqref{eq:one_loop_tmunu},
\begin{equation}
\mathcal T_{\mu \nu}^{(1)} 
= \frac{-i Q^2}{\kappa} 
\int \frac{d^d \boldsymbol{k}}{(2\pi)^d} 
\frac{V_{\mu \nu|0|0}(\boldsymbol{q}, \boldsymbol{k}, -(\boldsymbol{q} + \boldsymbol{k}))}
{\boldsymbol{k}^2 (\boldsymbol{k} + \boldsymbol{q})^2}
\label{eq:one_loop_RF2},
\end{equation}
we just have to replace the vertex $V$ with the corresponding graviton-photon-photon three-point vertex that we obtain from the new coupling terms. We will analyze all possible couplings one by one.
\subsection{Vertex $1: R_{\mu \nu \rho \sigma } F^{\mu  \nu } F^{\rho  \sigma}$}
First we calculate the vertex. Start with
\begin{multline}
R_{\mu\nu\rho\sigma}|_{\text{linearized}} F^{\mu\nu} F^{\rho\sigma} = 4 R_{\mu\nu\rho\sigma} |_{\text{linearized}}\eta^{\nu \alpha} \partial^\mu A_\alpha \eta^{\sigma \beta} \partial^\rho A_\beta \\= 2 \kappa \eta^{\nu \alpha} (\partial^\mu A_\alpha) \eta^{\sigma \beta} (\partial^\rho A_\beta) \left( \eta_{\lambda \nu} \eta_{\tau \rho} \partial_\mu \partial_\sigma + \eta_{\lambda \mu} \eta_{\tau \sigma} \partial_\nu \partial_\rho - \eta_{\lambda \nu} \eta_{\tau \sigma} \partial_\mu \partial_\rho - \eta_{\lambda \mu} \eta_{\tau \rho} \partial_\nu \partial_\sigma \right) h^{\lambda \tau}\,,
\label{eq:RF2_vertex}
\end{multline}
and after Fourier transforming we read off
\begin{equation}
(V_1)_{\mu \nu|\alpha|\beta}(q, k_1, k_2) = 4 i \lambda_1 \kappa\, k_1^\rho k_2^\sigma \left( \eta_{\mu \alpha} \eta_{\nu \sigma} q_\rho q_\beta + \eta_{\mu \rho} \eta_{\nu \beta} q_\alpha q_\sigma - \eta_{\mu \alpha} \eta_{\nu \beta} q_\rho q_\sigma - \eta_{\mu \rho} \eta_{\nu \sigma} q_\alpha q_\beta \right)\,.
\label{eq:RF2_vertex_fourier}
\end{equation}
We can then set $\alpha = \beta = 0$,
\begin{equation}
V_{\mu \nu|0|0}(\boldsymbol{q}, \boldsymbol{k}, -(\boldsymbol{q} + \boldsymbol{k})) = 4 i \kappa \lambda_1\, \eta_{0 \nu} \eta_{0 \mu} (\boldsymbol{k} \cdot \boldsymbol{q})(\boldsymbol{k} \cdot \boldsymbol{q} + \boldsymbol{q}^2)\,,
\label{eq:RF2_vertex_00}
\end{equation}
which leads to
\begin{equation}
 \mathcal T_{\mu \nu}^{(1)}  = 4 Q^2 \lambda_1\, \eta_{0 \nu} \eta_{0 \mu} \int \frac{d^d \boldsymbol{k}}{(2\pi)^d} \frac{(\boldsymbol{k} \cdot \boldsymbol{q})(\boldsymbol{k} \cdot \boldsymbol{q} + \boldsymbol{q}^2)}{\boldsymbol{k}^2 (\boldsymbol{k} + \boldsymbol{q})^2}\,.
\label{eq:RF2_Tmn_intermediate}
\end{equation}
Using
\begin{equation}
\boldsymbol{k} \cdot \boldsymbol{q} = \frac{1}{2} \left( (\boldsymbol{k} + \boldsymbol{q})^2 - \boldsymbol{k}^2 - \boldsymbol{q}^2 \right)
\label{eq:dot_product_trick}
\end{equation}
and setting scaleless integrals to zero,
\begin{equation}
 \mathcal T_{\mu \nu}^{(1)}  = -{Q^2 \lambda_1} \eta_{0 \nu} \eta_{0 \mu} \int \frac{d^d \boldsymbol{k}}{(2\pi)^d} \frac{\boldsymbol{q}^4}{\boldsymbol{k}^2 (\boldsymbol{k} + \boldsymbol{q})^2} = -{Q^2 \lambda_1} \eta_{0 \nu} \eta_{0 \mu}  \vb{q}^4 J_{(1)}(\vb{q}^2)\,.
\label{eq:RF2_Tmn_final}
\end{equation}
From this, we can get the form factors,
\begin{equation}
C_1\left(\boldsymbol{q}^2\right) = -{Q^2 \lambda_1}  \vb{q}^4 J_{(1)}(\vb{q}^2), \qquad C_2\left(\boldsymbol{q}^2\right) = 0
\label{eq:formfactors_RF2}
\end{equation}
Then, the metric is,
\begin{equation}
\begin{aligned}
h_{\mu \nu}(\boldsymbol{x}) &= -\frac{\kappa}{2} \left( \delta_\mu^0 \delta_\nu^0 + \frac{\eta_{\mu \nu}}{d - 1} \right) \left( -{Q^2 \lambda_1} \Jfactor \right) \int \frac{d^d \boldsymbol{q}}{(2\pi)^d} \frac{e^{i \boldsymbol{q} \cdot \boldsymbol{x}}}{\boldsymbol{q}^2} |\boldsymbol{q}|^d \\
&= -\frac{\kappa}{2} \left( \delta_\mu^0 \delta_\nu^0 + \frac{\eta_{\mu \nu}}{d - 1} \right) \left( -{Q^2 \lambda_1} \Jfactor \right) \left( \frac{1}{(4\pi)^{d/2}} \frac{\Gamma\left( \frac{d - (2 - d)}{2} \right)}{\Gamma\left( \frac{2 - d}{2} \right)} \left( \frac{2}{r} \right)^{d - (2 - d)} \right)\,.
\end{aligned}
\label{eq:metric_RF2}
\end{equation}
Upon simplification we find
\begin{equation}
\kappa h_{\mu v}(\boldsymbol{x})=-\left(\delta_\mu{ }^0 \delta_v{ }^0+\frac{\eta_{\mu v}}{d-1}\right)(d-2)^2 \frac{\kappa^2 Q^2 \lambda_1}{16 \pi^{d-1}}\left(\frac{1}{r}\right)^{2 d-2}\,.
\end{equation}
In particular, for $d=3$,
\begin{equation}
\kappa h_{00}(\boldsymbol{x}) = -\frac{G Q^2 \lambda_1}{\pi r^4}, \qquad \kappa h_{i j}(\boldsymbol{x}) = -\frac{G Q^2 \lambda_1}{\pi r^4} \delta_{i j}\,.
\label{eq:metric_RF2_d3}
\end{equation}

At one loop, the electromagnetic off-shell current is obtained from 
\begin{equation}
\begin{aligned}
\mathcal J_\mu^{(1)} 
&= \frac{i}{2M} \left(\includegraphics[valign=c]{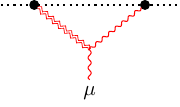}\right) \\
&= \frac{i}{2M} \int \frac{d^{d+1} k}{(2\pi)^{d+1}}
\left(-2 i Q M \eta^{\nu 0}\right)
\left(-i \kappa M^2 \eta^{\alpha 0} \eta^{\beta 0}\right)
\left(2\pi \delta(2 M k^0)\right)
\frac{
P_{\alpha \beta}^{\quad \alpha^{\prime} \beta^{\prime}} 
V_{\alpha^{\prime} \beta^{\prime}|\nu| \mu}(k, q, -(q+k))
}{
\left(i k^2\right)\left(i(k+q)^2\right)
} \\
&= \frac{i \kappa Q M}{2} \int \frac{d^d k}{(2\pi)^d}
\frac{
P_{00}^{ \quad \alpha^{\prime} \beta^{\prime}} 
V_{\alpha^{\prime} \beta^{\prime}|0|\mu}(\boldsymbol{k}, -(\boldsymbol{q}+\boldsymbol{k}), \boldsymbol{q})
}{
\boldsymbol{k}^2 (\boldsymbol{k}+\boldsymbol{q})^2
}\,,
\end{aligned}
\label{eq:Jalpha}
\end{equation}
where
\begin{equation}
P_{\mu \nu \mid \mu^{\prime} \nu^{\prime}} = \frac{1}{2}
\left(
\eta_{\mu \mu^{\prime}} \eta_{\nu \nu^{\prime}}
+ \eta_{\mu \nu^{\prime}} \eta_{\nu \mu^{\prime}}
- \frac{2}{d-1} \eta_{\mu \nu} \eta_{\mu^{\prime} \nu^{\prime}}
\right)\,.
\end{equation}
We can then get
\begin{equation}
\left(V_1\right)_{\mu \nu|0| 0} P^{\mu \nu}_{\quad 00}
= 4 \lambda_1 \left( \frac{d-2}{d-1} \right) i \kappa 
\big( (\boldsymbol{q}+\boldsymbol{k}) \cdot \boldsymbol{k} \big)
\boldsymbol{q} \cdot \boldsymbol{k}\,.
\end{equation}
Therefore,
\begin{align}
\mathcal J_0 
&= -2 \kappa^2 Q \lambda_1 M \left( \frac{d-2}{d-1} \right) 
\int \frac{d^d k}{(2\pi)^d} 
\frac{ \big( (\boldsymbol{q}+\boldsymbol{k}) \cdot \boldsymbol{k} \big)
(\boldsymbol{q} \cdot \boldsymbol{k}) }
{ \boldsymbol{k}^2 (\boldsymbol{k}+\boldsymbol{q})^2 } \nonumber \\
&= -\frac{1}{2} \kappa^2 Q \lambda_1 M \left( \frac{d-2}{d-1} \right) 
\int \frac{d^d k}{(2\pi)^d} 
\frac{ \boldsymbol{q}^4 }{ \boldsymbol{k}^2 (\boldsymbol{k}+\boldsymbol{q})^2 } \nonumber \\
&= \left( -\frac{1}{2} \kappa^2 Q \lambda_1 M \left( \frac{d-2}{d-1} \right) \right) 
\Jfactor
|\boldsymbol{q}|^d\,,
\end{align}
and
\begin{equation}
A_0(\boldsymbol{x}) =
\left( -\frac{1}{2} \kappa^2 Q \lambda_1 M \left( \frac{d-2}{d-1} \right) \right)
\Jfactor
\left[
\frac{1}{ (4\pi)^{d/2} }
\frac{ \Gamma\left( \frac{d-(2-d)}{2} \right) }
{ \Gamma\left( \frac{2-d}{2} \right) }
\left( \frac{2}{r} \right)^{ d - (2-d) }
\right]
\end{equation}
For $d = 3$,
\begin{equation}
    A_0 = \frac{GQ \lambda_1 M}{\pi r^4}\,.
\end{equation}

\subsection{Vertex $2 : R_{\mathrm{\mu \nu}} F^{\mathrm{\mu \sigma}} F^\nu{ }_\sigma$}
We start with 
\begin{equation}
    \begin{gathered}
R_{\mu\nu}|_{\text{linearized}} F^{\mu\rho} F^\nu{}_\rho 
=\eta^{\sigma\lambda}\left(\eta^{\rho \alpha} \partial^\mu-\eta^{\mu \alpha} \partial^\rho\right) A_\alpha \cdot\left(\eta_\rho{}^\beta \partial^\nu-\eta^{\nu \beta} \partial_\rho\right) A_\beta \\\quad  \frac{1}{2}\left(\eta_{\tau \mu} \eta_{\xi \lambda} \partial_\sigma \partial_\nu+\eta_{\tau \sigma} \eta_{\xi \nu} \partial_\mu \partial_\lambda-\eta_{\tau \mu} \eta_{\xi \nu} \partial_\sigma \partial_\lambda-\eta_{\tau \sigma} \eta_{\xi \lambda} \partial_\mu \partial_\nu\right) h^{\tau \xi}\,,
\end{gathered}
\end{equation}
and after Fourier transforming we obtain the cubic vertex:
\begin{multline}
\left(V_2\right)_{\tau \xi|\alpha| \beta}\left(q, k_1, k_2\right)=i \lambda_2 \eta^{\sigma\lambda}\left(\eta^{\rho}_{\alpha} k_1^\mu-\eta^{\mu}{ \alpha} k_1^\rho\right)\left(\eta_{\rho\beta} k_2^\nu-\eta^{\nu}_{\beta} k_{2\rho}\right)\\\left(\eta_{\tau \mu} \eta_{\xi \lambda} q_\sigma q_\nu+\eta_{\tau \sigma} \eta_{\xi \nu} q_\mu q_\lambda-\eta_{\tau \mu} \eta_{\xi \nu} q_\sigma q_\lambda-\eta_{\tau \sigma} \eta_{\xi \lambda} q_\mu q_\nu\right)\,.
\end{multline}
We then have
\begin{equation}
\begin{aligned}
V_{\mu \nu|0| 0}(\boldsymbol{q}, \boldsymbol{k},-(\boldsymbol{q}+\boldsymbol{k})) =& i \kappa \lambda_2 \Big( g_{\mu \nu}(\boldsymbol{k} \cdot \boldsymbol{q})^2 - \boldsymbol{q}^2 \boldsymbol{k}_\mu \boldsymbol{k}_\nu + \boldsymbol{k}^2 \boldsymbol{q}^2 g_{0 \nu} g_{0 \mu} \\ &+ \boldsymbol{k}_\nu \boldsymbol{q}_\mu(\boldsymbol{k} \cdot \boldsymbol{q}) + \boldsymbol{k}_\mu \boldsymbol{q}_\nu(\boldsymbol{k} \cdot \boldsymbol{q}) \\ &+ \boldsymbol{q}_\mu \boldsymbol{q}_\nu(\boldsymbol{k} \cdot \boldsymbol{q}) + \boldsymbol{q}^2 g_{0 \nu} g_{0 \mu}(\boldsymbol{k} \cdot \boldsymbol{q}) + \boldsymbol{q}^2 g_{\mu \nu}(\boldsymbol{k} \cdot \boldsymbol{q}) \Big)\,,
\end{aligned}
\end{equation}
We can then evaluate,
\begin{equation}
\mathcal T_{00}^{(1)} = Q^2 \lambda_2  \int \frac{d^d \boldsymbol{k}}{(2 \pi)^d} \frac{(\boldsymbol{k} \cdot \boldsymbol{q})^2 + \boldsymbol{k}^2 \boldsymbol{q}^2 + 2 \boldsymbol{q}^2(\boldsymbol{k} \cdot \boldsymbol{q})}{\boldsymbol{k}^2(\boldsymbol{k}+\boldsymbol{q})^2} = -Q^2 \lambda_2 \frac{3}{4} \int \frac{d^d \boldsymbol{k}}{(2 \pi)^d} \frac{\boldsymbol{q}^4}{\boldsymbol{k}^2(\boldsymbol{k}+\boldsymbol{q})^2}\,,
\end{equation}
\begin{equation}
\delta_{ij} \mathcal T^{ij}{}^{(1)} = Q^2 \lambda_2 \int \frac{d^d \boldsymbol{k}}{(2 \pi)^d} \frac{(2-d)(\boldsymbol{k} \cdot \boldsymbol{q})^2 - \boldsymbol{q}^2 \boldsymbol{k}^2 + (1-d)\boldsymbol{q}^2(\boldsymbol{k} \cdot \boldsymbol{q})}{\boldsymbol{k}^2(\boldsymbol{k}+\boldsymbol{q})^2} = Q^2 \lambda_2 \frac{d}{4} \int \frac{d^d \boldsymbol{k}}{(2 \pi)^d} \frac{\boldsymbol{q}^4}{\boldsymbol{k}^2(\boldsymbol{k}+\boldsymbol{q})^2}\,,
\end{equation}
which leads to the form factors
\begin{equation}
C_1\left(\boldsymbol{q}^2\right) = \frac{Q^2 \lambda_2(-2 d + 3)}{4(d - 1)} \Jfactor |\boldsymbol{q}|^d, \quad C_2\left(\boldsymbol{q}^2\right) = \frac{Q^2 \lambda_2 d}{4(d - 1)} \Jfactor |\boldsymbol{q}|^d\,,
\end{equation}
and, with $d=3$, the metric becomes
\begin{equation}
\kappa h_{00}=0, \quad \kappa h_{i j}=-\frac{3 G Q^2 \lambda_2}{\pi r^4} \frac{\vb x_i \vb x_j}{r^2}.
\end{equation}
For the vector potential, we first calculate,
\begin{equation}
\begin{aligned}
\left(V_2\right)_{\mu \nu|0|0}(\vb{k},\vb{q},-(\vb{k}+\vb{q})) P_{\mu \nu \mid 00}
= i \kappa \lambda_2\, 2 \Big[
&\,
(\vb{k}\!\cdot\!\vb{q})^2
+ \vb{k}^2 \vb{q}^2
+ 2 \vb{k}^2 (\vb{k}\!\cdot\!\vb{q}) \\
&\,
-\frac{1}{d-1}\!\left(
d(\vb{k}\!\cdot\!\vb{q})^2
+ 2 \vb{k}^2 \vb{q}^2
- 2(\vb{k}\!\cdot\!\vb{q})^2
+ d\,\vb{k}^2 (\vb{k}\!\cdot\!\vb{q})
\right)
\Big]\,.
\end{aligned}
\end{equation}
After dropping the terms which lead to scaleless integrals, what is left cancels. We conclude that the $\lambda_2$ higher derivative correction does not contribute to the one-loop order of $A_0$.

\subsection{Vertex $3: {R} {F}^{\mathrm{\mu \nu }} {F}_{\mathrm{\mu \nu }}$}
We begin once more by expanding 
\begin{equation}
\begin{aligned}
R F^{\rho\sigma} F_{\rho\sigma}|_{\text{linearized}} 
&= \eta^{\mu\nu} \eta^{\lambda\tau} \frac{1}{2} \left( \eta_{\mu' \lambda} \eta_{\nu' \nu} \partial_\mu \partial_\tau + \eta_{\mu' \mu} \eta_{\nu' \tau} \partial_\lambda \partial_\nu \right. \\
&\quad \left. - \eta_{\mu' \lambda} \eta_{\nu' \tau} \partial_\mu \partial_\nu - \eta_{\mu' \mu} \eta_{\nu' \nu} \partial_\lambda \partial_\tau \right) h^{\mu' \nu'} \cdot \left(2 \eta^{\alpha \sigma} \partial^\rho A_\alpha\right) \cdot \left(\eta_\sigma{}^\beta \partial_\rho - \eta_\rho{}^\beta \partial_\sigma \right) A_\beta\,,
\end{aligned}
\end{equation}
The cubic photon-photon-graviton vertex is
\begin{equation}
\left(V_3\right)_{\mu \nu \mid \alpha \beta}(q, k_1, k_2) = 4 \lambda_3 i \left( \eta_{\alpha \beta}(k_1 \cdot k_2) - k_{1\beta} k_{2\alpha} \right) \left( q_\mu q_\nu - \eta_{\mu \nu} q^2 \right)\,,
\end{equation}
In particular,
\begin{equation}
\left(V_3\right)_{\mu \nu|0|0}(\boldsymbol{q}, \boldsymbol{k}, -(\boldsymbol{q}+\boldsymbol{k})) = 4\kappa \lambda_3  i\, \boldsymbol{k} \cdot (\boldsymbol{q}+\boldsymbol{k}) \left( \boldsymbol{q}_\mu \boldsymbol{q}_\nu - \eta_{\mu \nu} \boldsymbol{q}^2 \right)\,.
\end{equation}
We can therefore find,
\begin{equation}
\mathcal T_{\mu \nu}^{(1)} = 4Q^2 \lambda_3 \int \frac{d^d \boldsymbol{k}}{(2\pi)^d} \frac{ \boldsymbol{k} \cdot (\boldsymbol{q}+\boldsymbol{k}) \left( \boldsymbol{q}_\mu \boldsymbol{q}_\nu - \eta_{\mu \nu} \boldsymbol{q}^2 \right) }{ \boldsymbol{k}^2 (\boldsymbol{k}+\boldsymbol{q})^2 }\,.
\end{equation}
From the expressions
\begin{equation}
\begin{aligned}
\mathcal T_{00}^{(1)} &= 4Q^2 \lambda_3 \int \frac{d^d \boldsymbol{k}}{(2\pi)^d} \frac{ \boldsymbol{k} \cdot (\boldsymbol{q}+\boldsymbol{k})\, \boldsymbol{q}^2 }{ \boldsymbol{k}^2 (\boldsymbol{k}+\boldsymbol{q})^2 } = -2{Q^2 \lambda_3} \int \frac{d^d \boldsymbol{k}}{(2\pi)^d} \frac{ \boldsymbol{q}^4 }{ \boldsymbol{k}^2 (\boldsymbol{k}+\boldsymbol{q})^2 }\,, \\
\delta^{ij} T_{ij} &= 4Q^2 \lambda_3 \int \frac{d^d \boldsymbol{k}}{(2\pi)^d} \frac{ (1-d)\, \boldsymbol{k} \cdot (\boldsymbol{q}+\boldsymbol{k})\, \boldsymbol{q}^2 }{ \boldsymbol{k}^2 (\boldsymbol{k}+\boldsymbol{q})^2 } = -2{Q^2 \lambda_3}(1-d) \int \frac{d^d \boldsymbol{k}}{(2\pi)^d} \frac{ \boldsymbol{q}^4 }{ \boldsymbol{k}^2 (\boldsymbol{k}+\boldsymbol{q})^2 }\,,
\end{aligned}
\end{equation}
we extract the form factors 
\begin{equation}
C_1\left(\boldsymbol{q}^2\right)=0, \quad C_2\left(\boldsymbol{q}^2\right)=2{Q^2 \lambda_3} \Jfactor |\boldsymbol{q}|^d\,.
\end{equation}
We can then compute the metric,
\begin{equation}
\kappa h_{00}=\frac{2 G Q^2 \lambda_3}{\pi r^4}, \quad \kappa h_{i j}=\frac{2 G Q^2 \lambda_3}{\pi r^4} \delta_{i j}-\frac{16 G Q^2 \lambda_3}{\pi r^4} \frac{\vb x_i \vb x_j}{r^2}\,.
\end{equation}
For the off-shell $U(1)$ current we need 
\begin{equation}
\begin{aligned}
\begin{gathered}
\left(V_3\right)_{\mu \nu|0|0} P^{\mu \nu}_{\quad 00}
= 4\lambda_3 \kappa i\,(\vb{q}\!\cdot\!(\vb{q}+\vb{k}))\!
\left[
\vb{k}_0^2 - \eta_{00}\vb{k}^2
- \frac{1}{d-1}\left(-\vb{k}^2 - d\,\vb{k}^2\right)
\right] \\[4pt]
= \frac{2 \lambda_3 \kappa i}{d-1}\,(\vb{q}\!\cdot\!(\vb{q}+\vb{k}))\,\vb{k}^2\,.
\end{gathered}
\end{aligned}
\end{equation}
This vertex leads to scaleless integrals and therefore does not contribute to \( A_0 \).

\subsection{Vertex 4: $\lambda_4 (\nabla \cdot F)^2$}

This vertex does not plays a role to leading order in $\lambda_4$. The reason is that the variation of this term in the action is proportional to the zeroth order equation of motion $\nabla_\mu F^{\mu\nu}=0$ (ignoring the pointlike charge Dirac delta source on the right hand side), and therefore it cannot affect the leading order correction to the equation of motion. 
Alternatively, $(\nabla \cdot F)^2$ can be removed to leading order by a field redefinition. Since under the integral sign it is equivalent to $-F^{\mu\nu}\nabla_\mu (\nabla\cdot F)_\nu$ it suffices to redefine $A_\nu\longrightarrow A_\nu - \lambda_4 (\nabla\cdot F)_\nu$ to remove it from the action.
In conclusion, there will be no change to the metric or gauge field to leading (linear) order in $\lambda_4$.

\subsection{Summary of Results}
 In summary, to leading order in the higher-derivative $RF^2$ perturbations, and working in the de Donder gauge, we found the following metric:
\begin{align}
    \kappa h_{00}(\boldsymbol{x}) &= \kappa h_{00}^{RN} + (2\lambda_3 -\lambda_1 )\frac{G Q^2}{\pi r^4} \nn
    \kappa h_{i j}(\boldsymbol{x}) &= \kappa h_{i j}^{RN} + \bigg[(2\lambda_3 - \lambda_1) \delta_{i j}- (16 \lambda_3 + 3 \lambda_2)  \frac{\boldsymbol{x}_i \boldsymbol{x}_j}{r^2}\bigg] \frac{G Q^2}{\pi r^4}\label{fin1}
\end{align}
and Maxwell field:
\begin{equation}
    A_0 = \frac{Q}{4 \pi r} + \lambda_1 \frac{G Q M}{\pi r^4}.\label{fin2}
\end{equation}
Higher loops will contribute to higher-order terms in the post-Minkowskian expansion, and higher order in the charge. Resumming the QFT loop expansion to arrive at a closed form for the metric is non-trivial. See for example \cite{Mougiakakos:2024nku}. 
 Even if the metric is known to all loop orders, the resummation is made difficult by the de Donder gauge. 
In the next section we will verify the results given in \eqref{fin1} and \eqref{fin2} by directly solving the modified Einstein-Maxwell equations, and working to all orders in $G$.

\section{Solving the Field Equations for the Perturbed  Black Hole Solution}
\label{sec:GR_method}

In this section we provide an alternative derivation of the black hole solution by explicitly solving field equations.  Varying the action in Eq.~\eqref{genhd1} with each field gives the field equations. We focus on terms with $\lambda_i$ and omit the $\lambda_4$ term as such a correction to the field equations vanishes within the perturbative analysis, to leading order. 
%
%
The variation with $A^\alpha$ gives the modified Maxwell equations:
\begin{equation}
\label{eq:field_eq_vector}
    \nabla_{\beta}\left( F^{\alpha\beta} - 4\lambda_1 R^{\alpha \beta}{}_{\mu\nu} F^{\mu\nu} - 4 \lambda_2  R^{[\beta}{}_{\mu} F^{\alpha] \mu} - 4 \lambda_3 R F^{\alpha \beta} - 4 \lambda_4 \nabla^{[\alpha}\nabla_{\mu}F^{\beta] \mu} \right)=0
\end{equation}
On the other hand, the variation with the inverse metric $g^{\alpha\beta}$ gives the modified Einstein equations:
\begin{equation}
\label{eq:field_eq_metric}
    G_{\alpha \beta} = 8\pi G \left(T_{\alpha\beta}+ \sum_{i=1}^3 \lambda_i E_{\alpha\beta}^{(i)}\right)\,,
\end{equation}
where $T_{\alpha\beta}$ is the stress-energy tensor for the vector field given by
\begin{equation}
    T_{\alpha\beta} =  F_{\alpha}{}^{\mu} F_{\beta \mu} - \frac{1}{4} F_{\mu\nu} F^{\mu\nu} g_{\alpha\beta}\,,
\end{equation}
while $E_{\alpha\beta}^{(i)}$ are given by
\begin{align}
    E_{\alpha\beta}^{(1)} = &  - 8 F_{(\alpha}{}^{\mu} F^{\nu\rho} R_{\beta)\mu \nu 
    \rho} + 2 F^{\mu\nu}F_{\alpha}{}^{\rho} R_{\beta 
    \rho \mu\nu} + \
g_{\alpha\beta} F^{\mu\nu}F^{\rho\sigma} R_{\mu\nu \rho\sigma} - 4 \nabla_{\nu}\nabla_{\mu}F_{(\alpha}{}^\mu F_{\beta)}{}^\nu\,, \\ 
    E_{\alpha\beta}^{(2)} = & 4 R_{\nu(\alpha} F_{\beta)}{}^{\mu} F_{\mu}{}^{\nu} -  2R_{\mu\nu}F_{\alpha}{}^{\mu}F_{\beta}{}^{\nu}  +  g_{\alpha\beta} R_{\mu\nu} F^{\mu\rho}F^{\nu}{}_{\rho}    + 2\nabla_{\mu}\nabla_{(\alpha} ( F_{\beta) \nu} F^{\mu \nu}) \nonumber \\
    &-  \Box (F_{\alpha\mu}F_{\beta}{}^\mu) \
-  g_{\alpha\beta} \nabla_{\nu}\nabla_{\mu}(F^{\mu\rho}F^{\nu}{}_{\rho})\,, \\ 
    E_{\alpha\beta}^{(3)} = & -2[G_{\alpha\beta} F_{\mu\nu}F^{\mu\nu} + 2 R F_{\alpha}{}^{\rho} F_{\beta \rho} -  \nabla_{\beta}\nabla_{\alpha}(F_{\mu\nu}F^{\mu\nu}) + g_{\alpha\beta}  \Box (F_{\mu\nu}F^{\mu\nu})]\,.
    %
\end{align}

We now derive a spherically symmetric and static black hole solution by solving the above field equations. We work in small coupling  approximation and keep only to linear order. However, unlike the QFT approach, we do not work in post-Minkowskian approximation.  We first work in Schwarzschild-like spherical coordinates $(t,r,\theta,\phi)$, and next perform a coordinate transformation to move to de Donder gauge so that we can compare the result in this section to that from the QFT calculation.


We start by considering the following metric and vector field ansatz:
\begin{align}
\label{eq:metric_ans}
ds^2 = & - N(r)^2 f(r) dt^2 + \frac{dr^2}{f(r)} + r^2 (d\theta^2 + \sin^2 \theta d\phi^2)\,, \\
\label{eq:vector_ans}
    A_\alpha dx^\alpha = & \left(\frac{Q}{4\pi\, r} + \epsilon \, \delta a(r) \right)dt\,,
\end{align}
with 
\begin{align}
    N(r) =& 1 + \epsilon \, \delta N(r)\,, \\
    f(r)=&  1 - \frac{2GM}{r} + \frac{GQ^2}{4 \pi r^2} + \epsilon \, \delta f(r)\,, \label{eq:metric_f}
\end{align}
where $M$ is the black hole mass, $Q$ represents the black hole charge, and $\epsilon$ is the book-keeping parameter to count the order of the coupling constants $\lambda_i$. 
The absence of $A_r$ is due to the Lorenz gauge.

We plug the above ansatz into the field equations, in which we replace $\lambda_i$ by $\epsilon \lambda_i$. 
We treat the non-GR correction perturbatively by assuming $\epsilon \ll 1$. 
We expand the field equations about $\epsilon = 0$.
In this case, $\mathcal{O}(\epsilon^0)$ equations are automatically satisfied because of the ansatz. At $\mathcal{O}(\epsilon^1)$, the $t$-component of the modified Maxwell equations yields
\begin{align}
r \delta a''+2 \delta a'+\frac{Q}{4\pi r} \delta N' = \frac{GQ}{\pi^2  r^6} 
   \left[12 \pi 
   \lambda_1 M r-Q^2 (6 \lambda_1+\lambda_2)\right]\,,
\end{align} 
while the $(t,t)$ and $(r,r)$ components of the modified Einstein equations give
\begin{align}
   2 \pi  r^2 \left(r
   \delta f'-2 G Q \delta a'+\delta f \right)-G Q^2 \delta N 
   =&S_1\,, \\
   G Q^2 \delta N-4\pi r \Delta \delta N' -2 \pi  r^2 \left(r \delta f'-2 G Q
   \delta a'+\delta f\right) =&S_2
\end{align}
where
\begin{align}
    \Delta =& r^2-2 G 
   M r +\frac{GQ^2}{4\pi}\\
    S_1 =& \frac{G Q^2}{ 2 \pi  r^4} \left\{12 \pi
    r^2 (4 \lambda_1+3 \lambda_2+8 \lambda_3)-16 \pi  G M r
   [5 (\lambda_1+\lambda_2)+14 \lambda_3]+G Q^2
   (4 \lambda_1+9 \lambda_2+30 \lambda_3)\right\}\,,\\
    S_2 = & \frac{G Q^2}{2 \pi  r^4}
   \left[4 \pi  r^2 (3 \lambda_2+16 \lambda_3)-16 \pi  G M r (\lambda_1+\lambda_2+6
   \lambda_3)+G Q^2 (8 \lambda_1+3 \lambda_2+10
   \lambda_3)\right]\,.
\end{align}

We solve these equations under the following boundary condition at infinity:
\begin{equation}
\label{Asympt}
    \delta a = \mathcal{O}\left(\frac{1}{r^2}\right)\,, \quad  \delta g_{tt} = \mathcal{O}\left(\frac{1}{r^2}\right)\,,
\end{equation}
where $\delta g_{tt}$ is the $\mathcal{O}(\epsilon)$ part of $g_{tt}$. The constant terms in $\delta a$ and $\delta g_{tt}$ are absent to maintain asymptotic flatness while $1/r$ terms can be absorbed to $M$ and $\alpha$. We find the following solutions (these are exact, without expansion in large $r$):
\allowdisplaybreaks
\begin{align}
    \delta f=& -\frac{G Q^2}{20 \pi ^2 r^6} \left[20 \pi  (4 \lambda_1+3 \lambda_2+8 \lambda_3)r^2 -20 \pi (7 \lambda_1+5 \lambda_2+14 \lambda_3) G M r \right. \nonumber \\
    &\left.+(16 \lambda_1+11
   \lambda_2+30 \lambda_3)G Q^2 \right]\,,\\
    \delta N = & \frac{G Q^2}{2 \pi  r^4} [3 (\lambda_1+\lambda_2)+10 \lambda_3]\,,\\
    \delta a =& \frac{G Q}{40 \pi ^2 r^5} \left[40 \pi  \lambda_1 M r+ (-9 \lambda_1+\lambda_2+10 \lambda_3)Q^2\right]\,.
\label{electricpot}
\end{align}
We have checked that the above solution for $\delta f$ matches with that in~\cite{KatsMotlPadi2007} ($\delta N$ and $\delta a$ are not presented in the reference) while all of $\delta f$, $\delta N$ and $\delta a$ match with the one in~\cite{delRio:2024kms} for the DH case.


In order to compare the above results with those derived from QFT, we will now move to de Donder gauge with the condition
\begin{equation}
\label{eq:de_Donder}
    \eta^{\mu\nu}\Gamma^\alpha_{\mu\nu} = 0\,,
\end{equation}
where $\eta^{\mu\nu}$ is (the inverse of) the Minkowski spacetime while $\Gamma^\alpha_{\mu\nu}$ are the Christoffel symbols. We will follow the procedure outlined in~\cite{DOnofrio:2022cvn}.

We begin by changing to Cartesian coordinates $(x,y,z)$ via the standard transformation
\begin{equation}
    (x,y,z) = (r \sin\theta \cos \phi, r \sin\theta \sin \phi, r \cos\theta)\,.
\end{equation}
The metric in Eq.~\eqref{eq:metric_ans} then becomes 
\begin{equation}
    ds^2 = - N(r)^2 f(r)dt^2 + d \vec x^2 + \frac{1 -f(r)}{f(r)} \frac{(\vec x \cdot d \vec x)^2}{r^2}\,.
\end{equation}
In order to satisfy the de Donder gauge condition, we further carry out a coordinate transformation
\begin{equation}
    (t,\vec x) \longrightarrow (t, F(r) \vec x)\,.
\end{equation}
 The line element in the new coordinates becomes
 \begin{equation}
    ds^2 = -h_0(r) dt^2 + h_1(r) d\vec x^2 + h_2 \frac{(\vec x \cdot d \vec x)^2}{r^2}\,,
 \end{equation}
 with
 \begin{align}
     h_0(r) = & f(F(r) r)\,, \\
     h_1(r) =& F(r)^2\,, \\
     h_2(r) = & -F(r)^2 + \frac{[F(r)+r F'(r)]^2}{f(F(r) r)}\,.
 \end{align}
 The de Donder gauge condition in Eq.~\eqref{eq:de_Donder} then yields
 \begin{equation}
 \label{eq:de_Donder_2}
     h_0'+h_1'-h_2' = \frac{4}{r} h_2\,.
 \end{equation}

 Let us now derive the metric functions $h_i$ perturbatively about $r=\infty$. One can first solve Eq.~\eqref{eq:de_Donder_2} perturbatively to find $F(r)$:
  \begin{align}
     F(r) =& 1+\frac{G M}{r}+\frac{2 G^2 M^2}{r^2}-\frac{2 G^2 M \left(G
   M^2-\tfrac{Q^2}{2 \pi} \right)}{3
   r^3} \log \left(\frac{2 G M}{r}\right) \nonumber \\
   &+\left[ G^3 M^2 \left(4 G
   M^2+\tfrac{7Q^2}{4 \pi} \right)+8G^3 M^2 \left(G M^2-\tfrac{Q^2}{2 \pi} \right)
   \log \left(\frac{2 G M}{r}\right)\right]\frac{1}{6r^4} \nonumber \\
   &+\epsilon  \frac{2(\lambda_1-2 \lambda_3) G Q^2}{4 \pi r^4} +\mathcal{O}\left( \frac{1}{r^5} \right)\,.
 \end{align}
 Using this, $h_i(r)$ become 
 \begin{align}
     h_0(r) = & 1-\frac{2 G M}{r}+\frac{G \left(2 G
   M^2 + \tfrac{Q^2}{4 \pi}\right)}{r^2} +\frac{2 G^2 M \left(G M^2-\tfrac{Q^2}{4 \pi}
   \right)}{r^3} \nonumber \\
   &  - \left[\tfrac{3Q^2}{4 \pi}+18 G M^2 +4\left(G M^2-\tfrac{Q^2}{2 \pi}\right) \log \left(\frac{2 G
   M}{r}\right)\right] \frac{G^3 M^2}{3 r^4} \nonumber \\
   &  
   -\epsilon \frac{4G
   Q^2  \left(\lambda_1-2 \lambda_3\right)}{4 \pi r^4 } +\mathcal{O}\left( \frac{1}{r^5} \right)\,, \\
        h_1(r) =& 1 +\frac{2 G M}{r}+\frac{5 G^2
   M^2}{r^2}+ \left[3 G M^2-\left(G M^2-\tfrac{Q^2}{2 \pi}\right) \log
   \left(\frac{2 G M}{r}\right)\right]\frac{4 G^2 M}{3 r^3} \nonumber \\
&   + \left[16 G M^2+ \tfrac{7Q^2}{4 \pi} +4 \left(G M^2-\tfrac{Q^2}{2 \pi}\right) \log \left(\frac{2 G
   M}{r}\right)\right]\frac{G^3 M^2 }{3 r^4} \nonumber \\
&  +\epsilon\frac{G Q ^2  
   (\lambda_1-2 \lambda_3)}{ \pi r^4} +\mathcal{O}\left( \frac{1}{r^5} \right)\,,\\
        h_2(r) =& -\frac{G \left(\tfrac{Q^2}{4 \pi} +7 G
   M^2\right)}{r^2}
   -\left[19 G M^2+ \tfrac{7Q^2}{4 \pi} -6 \left(G M^2-\tfrac{Q^2}{2 \pi}\right) \log \left(\frac{2 G
   M}{r}\right)\right]\frac{2 G^2 M }{3 r^3} \nonumber \\
&-\left[58 G^2 M^4 + \tfrac{13Q^2}{4 \pi} G
   M^2-\tfrac{3Q^4}{{16 \pi^2}}+8 G
   M^2\left(G M^2-\tfrac{Q^2}{2 \pi}\right) \log \left(\frac{2 G M}{r}\right)\right]\frac{G^2}{3 r^4}\nonumber \\
&+\epsilon\frac{G Q^2 (3 \lambda_2+16 \lambda_3)}{\pi r^4} +\mathcal{O}\left( \frac{1}{r^5} \right)\,.
 \end{align}
 Notice that the $\mathcal{O}(\epsilon)$ terms in the above equations match with the expressions found in Eq.~\eqref{fin1} derived from the classical limit of  the QFT off-shell currents. 

The vector potential can be found similarly; in this case, we merely must transform the radial coordinate in the argument of the scalar field, yielding
\begin{equation}
    A_\alpha dx^\alpha =  \left(\frac{Q}{4\pi\, F(r) r} + \epsilon \, \delta a(F(r)r) \right)dt\,,
\end{equation}
where 
\begin{align}
   \frac{Q}{4\pi F(r)r}&=\frac{Q}{4\pi r}-\frac{GMQ}{4\pi r^2}-\frac{G^2M Q  M^2 }{4 \pi
    r^3}\nonumber\\
   &+\frac{Q \left[2 \pi  G^3 M^3 \log \left(\frac{2 G M}{r}\right)+9 \pi  G^3 M^3-G^2 M Q^2
   \log \left(\frac{2 G M}{r}\right)\right]}{12 \pi ^2 r^4}+\mathcal{O}\left( \frac{1}{r^5}\right) ,\\
     \delta a_0(F(r)r)&=\epsilon\frac{GQ\lambda_1 M}{\pi r^4} +\mathcal{O}\left( \frac{1}{r^5}\right).
\end{align}
The order $\mathcal O(\epsilon)/r^4$ term above matches 
the earlier QFT result in Eq.~\eqref{fin2}\footnote{Terms of order $\lambda GQ^3/r^5$ arise at two-loop and so do terms of order $\lambda G^2M/r^5$.}.

With this, we have computed the metric and vector potential exactly to linear order in the higher-derivative coupling constants, transformed them into de Donder gauge used for the field-theoretic method above, and showed agreement with the QFT results computed in Sect.~\ref{sec:BH_from_worldline}.

\section{Thermodynamics}
    \label{sec:thermodynamics}

In this section, we study the thermodynamic properties of the new black hole solutions. First, 
 from the boundary condition in Eq.~\eqref{Asympt},
it is clear that the mass and charge of the perturbed black holes do not change relative to the Reissner-Nordstrom values of $M$ and $Q$, respectively.

In terms of the Reissner-Nordstr\"om outer (+)/inner (-) horizons,
\begin{equation}
    r_{\pm} = G M \pm \sqrt{G}\sqrt{G M^2 -Q^2/(4\pi)}\,,\label{eq:rpm}
\end{equation}
 the outer horizon of the perturbed black holes (in the coordinate system  \eqref{eq:metric_ans}) is shifted 
to 
\begin{align}
r_{\text{outer\,H}}=r_{+}\,+
\frac{  G Q^2 \left[5 r_+^2 (\lambda_1+\lambda_2+2 \lambda_3)- G Q^2(3 \lambda_1+3 \lambda_2+10 \lambda_3)/(4\pi) \right]}{10\pi r_+^3[r_+^2-GQ^2/(4\pi)]},\label{eq:r_outer}
\end{align}
and, 
similarly, the inner horizon is shifted to
\begin{align}
r_{\text{inner\,H}}=r_{-}\,+
\frac{  G Q^2 \left[5 r_-^2 (\lambda_1+\lambda_2+2 \lambda_3)- G Q^2(3 \lambda_1+3 \lambda_2+10 \lambda_3)/(4\pi) \right]}{10\pi r_-^3[r_-^2-GQ^2/(4\pi)]},\label{eq:r_inner}
\end{align}
where we consistently keep only terms up to linear in the couplings $\lambda_i$.

Next we will derive the black hole entropy from the first law of black hole thermodynamics.
The RN black hole temperature is 
\begin{equation}
    T_0=\frac{r_+-r_-}{4\pi r_+^2}=\frac{r_+^2-(GQ^2)/(4\pi)}{4\pi r_+^3}
\end{equation}
and that of the perturbed RN black holes is obtained in terms of surface gravity from
\begin{equation}
    T=\frac{1}{4\pi} \frac{1}{N}\partial_r(fN^2)|_{r=r_{\text{outer\,H}}}.
\end{equation}
Since in the vicinity of the outer horizon $f\approx (r-{r_{\text{outer\,H}}})\frac{d}{dr}f(r_{\text{outer\,H}})$ this evaluates to
\begin{align}
    T=&\frac{1}{4\pi}N(r_{\text{outer\,H}})\frac{d}{dr}f(r_{\text{outer\,H}}) \nonumber \\
     =& T_0 -\frac{G Q^2 \left[(7 \lambda_1+2 \lambda_2)G^2 Q^4 -8 \pi (7 \lambda_1+2 \lambda_2)  G Q^2 r_+^2 +80 \pi ^2 \lambda_1 r_+^4\right]}{80 \pi ^3 r_+^7 \left(4 \pi  r_+^2-G Q^2\right)}\nn
     \equiv & T_0+\delta T.
 \end{align}
From the previous section, using \eqref{electricpot}, we find the 
 value of the electric potential at the horizon\footnote{Note that due to the non-minimal $RF^2$ couplings, the Maxwell field equation has changed. Therefore we cannot obtain $A_t $ by the usual Gauss formula
$\int d^3 \vec r \,\partial_r[\sqrt{-g} g^{rr}g^{tt} \partial_r A_t]=Q$. Instead, the right hand side of Maxwell's equations receives corrections from $RF^2$-terms. The result for the gauge potential is the one quoted in the text.}:\
\begin{align}
\Phi_{\text H}&=A_t(r_{\text{outer\,H}})=\frac{Q}{4\pi r_+}\bigg(1+\frac{(7 \lambda_1+2 \lambda_2)G^2 Q^4 -8 \pi (7 \lambda_1+2 \lambda_2)  G Q^2 r_+^2 +80 \pi ^2 \lambda_1 r_+^4}{10 \pi  r_+^4 \left(4 \pi  r_+^2-G Q^2\right)}\bigg)\nn
&=\Phi_{0}-\frac{\delta T}{2}\frac{4\pi r_+^2}{GQ}.
\end{align}
With these ingredients, we can now determine the black hole entropy. We notice that the perturbed black hole mass still depends only on two dimensional parameters, $r_+$ and $Q$:
\begin{equation}
    M=\frac{GQ^2+4\pi r_+^2}{8\pi G r_+}.
\end{equation}
We first check to see if the following first law \cite{BardeenCarterHawking1973} holds: 
\begin{equation}
    dM=T dS+\Phi_{\text H} dQ,
\end{equation}
with $dM=(\partial_{r_+} M )\,dr_+ +(\partial_Q M)\, dQ$.
This leads to 
\begin{equation}
(\partial_{r_+} S) \, dr_++ (\partial_Q S)\, dQ=\frac{1}T \bigg[(\partial_{r_+} M) \,dr_+ +(\partial_Q M-\Phi_{\text H})\, dQ\bigg].
\end{equation}
After a quick inspection, we verify that the right hand side is exact. Therefore we can extract the perturbed black hole entropy. Trading off $r_+$ in terms of the outer horizon area $A_\mathcal H=4\pi r_{\text{outer\,H}}^2$ gives the entropy. Working to leading order in the perturbations we find:
\begin{equation}
    S=\frac{A_\mathcal H}{4G}-\frac{4\pi (2\lambda_1+\lambda_2+2\lambda_3)Q^2 }{A_\mathcal H}.
\label{entropy}
\end{equation}
In Appendix \ref{AppendixIW} we give an alternative derivation of the black hole entropy using the definition given by Iyer and Wald \cite{Iyer:1994ys}. The outcome is the same as in Eq.\eqref{entropy}.

The deviation from the usual Bekenstein-Hawking entropy $A_{\mathcal H}/(4G)$ \cite{Bekenstein1973} \cite{Hawking1975} found in \eqref{entropy}  appears to be typical of higher-derivative perturbed black holes.
The case of a RN black hole perturbed by the $R^2$ terms (the $a_1\neq 0, a_2\neq 0$ terms in \eqref{genhd1}) was considered in \cite{Campanelli:1994jt, Campanelli:1994sj}. The identification with the parameters used in \cite{Campanelli:1994jt, Campanelli:1994sj} is $a_1=\alpha, a_2=\beta$. In this case, the entropy correction was found to be  $-8\pi^2 a_2 Q^2/A_\mathcal H$, similar to \eqref{entropy}. 

At extremality, when $r=r_{\text{extrem H}}$,  the horizon is degenerate:
$f(r_{\text{extrem H}})=0$ and $f'(r_{\text{extrem H}})=0$.
These conditions are satisfied for \begin{align}
r_{\text{extrem H}}=
\frac{\sqrt{G}Q}{\sqrt{4\pi}}+\frac{8\sqrt\pi\,\lambda_3}{\sqrt G Q}\label{r_ext}
\end{align}
when the mass is equal to 
\begin{align}
    M|_{\text{extremal}}=\frac{Q}{\sqrt {4\pi G}}\left[1-\frac {8\pi}{5 G Q^2}\left(\lambda_1+\lambda_2\right)\right]=M_0|_{\text{extremal}}+\delta M.
\label{eq:M_ext}
\end{align}
This agrees with \cite{KatsMotlPadi2007}\footnote{We need to be careful when comparing with \cite{KatsMotlPadi2007}. Naively we match $c_4=\lambda_3, c_5=\lambda_2, c_6=\lambda_1$. However, they also have $(\nabla F)^2$ terms which, we argued in Introduction, reduce to $R F^2$ terms using Bianchi identities. First we eliminated $c_8$ which changed $c_9$ to $c_9+2c_8$. Then we eliminated $c_9$. This changed $c_5$ and $c_6$. So the mapping is $\lambda_3=c_4$, $\lambda_2 = c_5-(c_9+2c_8)$ and $\lambda_1=c_6+\frac 12 (c_9+2c_8)$. This accounts for the full $c_4, c_5, c_6, c_8, c_9$ dependence in \cite{KatsMotlPadi2007}.} 
.
Weak gravity conjecture \cite{Arkani-Hamed:2006emk} would require that the correction factor $\delta M$ be negative, that is,
\begin{align}\lambda_1+\lambda_2>0. 
\label{wgc}\end{align} Of course, if there are additional higher-derivative corrections, they need to be accounted for as well.\footnote{Including all parameters, similar to \cite{KatsMotlPadi2007},  from the weak gravity conjecture one gets a condition on a particular linear combination of all the coupling constants in \eqref{genhd1}. It is amusing to note that knowledge of light-by-light scattering parameters, $b_{1,2}$, could then be used to place constraints on the couplings of the gravitational higher-derivative terms, $\lambda_i$ and $a_j$.}

 At the same time, we observe that even when $\lambda_1+\lambda_2>0$, the sign of the entropy change relative to the extremal Reissner-Nordstrom 
remains indeterminate. We have:
\begin{align}
S|_{\text{extremal}}&=
\frac {Q^2}4-\frac{4\pi(2\lambda_1+\lambda_2)}G\equiv S_0|_{\text{extremal}}+\delta S.\label{s_ext}
\end{align}
where the extremal RN entropy is $S_0|_{\text{extremal}}=Q^2/4$. The correction to the entropy, $\delta S$, receives contributions from the change in the position of the outer horizon \eqref{eq:r_outer} and the deviation from the Bekenstein-Hawking entropy given in \eqref{entropy}.  As noticed in \cite{McPeak:2021tvu}, the extremal black hole  corrections to mass \eqref{eq:M_ext}, entropy \eqref{s_ext} and horizon \eqref{r_ext} in the higher-derivative theory relative to the extremal RN black hole, are proportional to linearly independent combinations of $\lambda_i$. 

We can also compare the shift in the entropy for a solution with $M=M_0|_{\text{extremal}}=Q/\sqrt{4\pi G}$, which is the mass of the extremal RN black hole. 
In the presence of the higher-derivative corrections, the location of the outer horizon is shifted by terms of order $\sqrt{\lambda_i}$  to  
\begin{align}
r_{\text{outer/inner H}}=\frac{\sqrt G Q}{\sqrt {4\pi}}\pm\frac{2\sqrt{\lambda_1+\lambda_2}}{\sqrt{5}}+
\frac{4\sqrt \pi (\lambda_1+\lambda_2+10\lambda_3)}{5\sqrt G Q}.
\end{align}
Note that  the linear expansion in $\lambda_i$ we performed earlier in \eqref{eq:r_outer} to determine the location of the horizons is no longer valid for a mass equal to the extremal RN mass $(M_0|_{\text{extremal}}$). In particular \eqref{eq:r_outer} blows up when substituting $M=M_0|_{\text{extremal}}$. This aspect was first commented in \cite{Hamada:2018dde}.

Similarly, using \eqref{entropy}, the entropy of this solution  is 
\begin{align}
    S|_{M=M_0|_{\text{extremal}}}=S_0|_{\text{extremal}}+\frac{2Q\sqrt {\pi(\lambda_1+\lambda_2)}}{ \sqrt{5G}}-\frac{\pi (8\lambda_1+3\lambda_2)}{5G}.
\end{align}

As a result of carrying the appropriate expansion in $\sqrt{\lambda_i}$ for a mass $M=M_0|_{\text{extremal}}$, unlike in \cite{Cheung:2018cwt} where the expansion is in  $\lambda_i$, we find that the shift in entropy relative to the extremal RN black hole is not divergent. 
However, 
the statement  made by the authors \cite{Cheung:2018cwt} remains valid: if the weak gravity conjecture is valid, that is if $\lambda_1+\lambda_2>0$,  then the entropy shift is positive.

Our observation is that, since both the leading order entropy and horizon corrections are proportional to $\sqrt{\lambda_1+\lambda_2}$, unless this expression is real, the solution with $M=M_0|_{\text{extremal}}$ has a naked singularity.
So the weak gravity conjecture is reduced to the statement that $M=M_0|_{\text{extremal}}$ is a black hole and not a singular solution\footnote{More generally, if we consider a solution with $M=M_0|_{\text{extremal}}(1-\frac{8\pi}{5GQ^2} (a\lambda_1+b\lambda_2))$ with $a, b$ real parameters, the horizons are at $\sqrt {G /(4\pi)} \,Q\pm 2\sqrt{((1-a)\lambda_1+(1-b)\lambda_2)/5} $. The condition for the reality of the horizon is that $(1-a)\lambda_1+(1-b)\lambda_2>0$  which translates into $M-M|_{\text{extremal}}>0$. }. {\it And since for all black holes we must have $M>M|_{\text{extremal}}$, this implies $0>\delta M$ (and equivalently, a real horizon, a positive entropy shift $2Q\sqrt {\pi(\lambda_1+\lambda_2)/(5 G)}>0$ and a positive temperature shift $\delta T=4\sqrt{\lambda_1+\lambda_2}/(\sqrt 5 \, G Q^2)$), which is the weak gravity conjecture}.


There is one more conclusion to be drawn from requiring that the extremal RN mass solution remains a black hole in the presence of the higher-derivative corrections. Substituting the values  $\lambda_{1,2}$ for the DH Lagrangian in \eqref{dh}, we see that these coefficients, obtained by integrating out a spin 1/2 charged fermion coupled to both Maxwell and gravitational fields, violate 
the inequality  \eqref{wgc}.
This rules out DH theory\footnote{If instead we consider the coefficients determined by integrating out a spin 0 charged particle, we find that the inequality \eqref{wgc} remains unsatisfied, since both $\lambda_1$ and $\lambda_2$ are negative (see equation (5.5) in \cite{Bastianelli2009}).} as a stand-alone extension of Einstein-Maxwell theory. 
As for the Horndeski theory, using \eqref{wgc} we can fix the sign of the overall coefficient $\gamma$ in \eqref{horndeski}, namely $\gamma<0$.

 
\section{Observational Bounds from Black Hole Shadows}
\label{sec:BH_shadow}

In this section, we derive constraints on the $RF^2$ black hole from black hole shadow observations by the Event Horizon Telescope~\cite{EventHorizonTelescope:2019dse,EventHorizonTelescope:2022wkp}\footnote{{In $RF^2$ gravity, the equation of motion for a photon,  $g_{\mu\nu} k^\mu k^\nu = 0$ with the wave vector $k^\mu$, is modified with a source on the right hand side proportional to $\lambda_i$~\cite{Shore:2003zc,HertzbergNathanSemaan2025}. For simplicity, we ignore this correction when estimating constraints from black hole shadow observations and focus on the correction in the metric. Therefore, bounds obtained here should be taken as orders of magnitude estimate.}}. The shadow radius $r_\mathrm{sh}$ is related to the photon sphere radius $r_\mathrm{ph}$ by~\cite{Perlick:2021aok}
\begin{equation}
\label{eq:r_shadow}
r_\mathrm{sh} = \frac{r_\mathrm{ph}}{\sqrt{-g_{tt}(r_\mathrm{ph})}}\,,    
\end{equation}
where the photon sphere radius is obtained by solving the following equation~\cite{Perlick:2021aok}:
\begin{equation}
\label{eq:rph_eq}
    g_{tt}(r_\mathrm{ph}) - \frac{r_\mathrm{ph}}{2} g_{tt}'(r_\mathrm{ph}) =0\,.
\end{equation}
The shadow radius $r_\mathrm{sh}$ of Sgr A* has been measured with a 1-$\sigma$ error of~\cite{Vagnozzi:2022moj} 
\begin{equation}
    4.55 GM < r_\mathrm{sh} < 5.22GM\,.
    \label{eq:r_sh_bound}
\end{equation}
Vagnozzi \emph{et al}.~\cite{Vagnozzi:2022moj} used this bound to find a constraint on the electric charge for a RN black hole as $Q < 2.83 \sqrt{G} M$ (after accounting for the difference in the definition of $Q^2$ between this paper and Ref.~\cite{Vagnozzi:2022moj} by a factor of $4\pi$). 

Let us now apply the bound in Eq.~\eqref{eq:r_sh_bound} to constrain the $RF^2$ theory. First, solving Eq.~\eqref{eq:rph_eq}, we find
\begin{equation}
    r_\mathrm{ph} = r_0 + \frac{2 \left[  (23 \lambda_1+13 \lambda_2+30
   \lambda_3)\pi M(r_0-3  G M)+ (13 \lambda_1+8
   \lambda_2+10 \lambda_3)Q^2\right]}{5 \left[3 G M \left(6
   \pi  M r_0-Q^2\right)-2 Q^2 r_0\right]}\,,
\end{equation}
where $r_0$ is the photon sphere radius for a RN black hole given by
\begin{equation}
    r_0 = \frac{1}{2} \left(3 G M+\sqrt{G \left(9 G M^2-\frac{2 Q^2}{\pi}\right)}\right)\,.
\end{equation}
Notice that $r_0$ exists only when $|Q| \leq 3\sqrt{\pi G/2} M \approx 3.76 \sqrt{G} M$.
Next, by substituting the above $r_\mathrm{ph}$ to Eq.~\eqref{eq:r_shadow}, we find $r_\mathrm{sh}$ as
\begin{align}
    r_\mathrm{sh} = &\frac{2 \sqrt{\pi} r_0^2}{\sqrt{G \left(4 \pi M r_0-Q^2\right)}} -\frac{\sqrt{G} Q^2 }{5
   \sqrt{\pi } r_0^2 \left(4 \pi  M r_0-Q^2\right)^{3/2}
   \left[3 G M \left(Q^2-6 \pi  M r_0\right)+2 Q^2
   r_0\right]} \nonumber \\ 
   &\times \left\{180 \pi (2 \lambda_1+\lambda_2)  G^2 M^3  \left(6 \pi  M r_0-Q^2\right) \right.  \nonumber \\ 
   & \left. +G M Q^2
   \left[(67 \lambda_1+32 \lambda_2)Q^2 -18 \pi (29 \lambda_1+14 \lambda_2)  M
   r_0 \right]+2 (9 \lambda_1+4 \lambda_2) Q^4
   r_0 \right\}\,.
\end{align}
The above $r_\mathrm{sh}$ reduces to $r_\mathrm{sh} = 3\sqrt{3}GM$ in the limit $Q \to 0$. Notice that the $RF^2$ correction to $r_\mathrm{sh}$ only depends on $\lambda_1$ and $\lambda_2$.

\begin{figure}
\includegraphics[width=9.5cm]{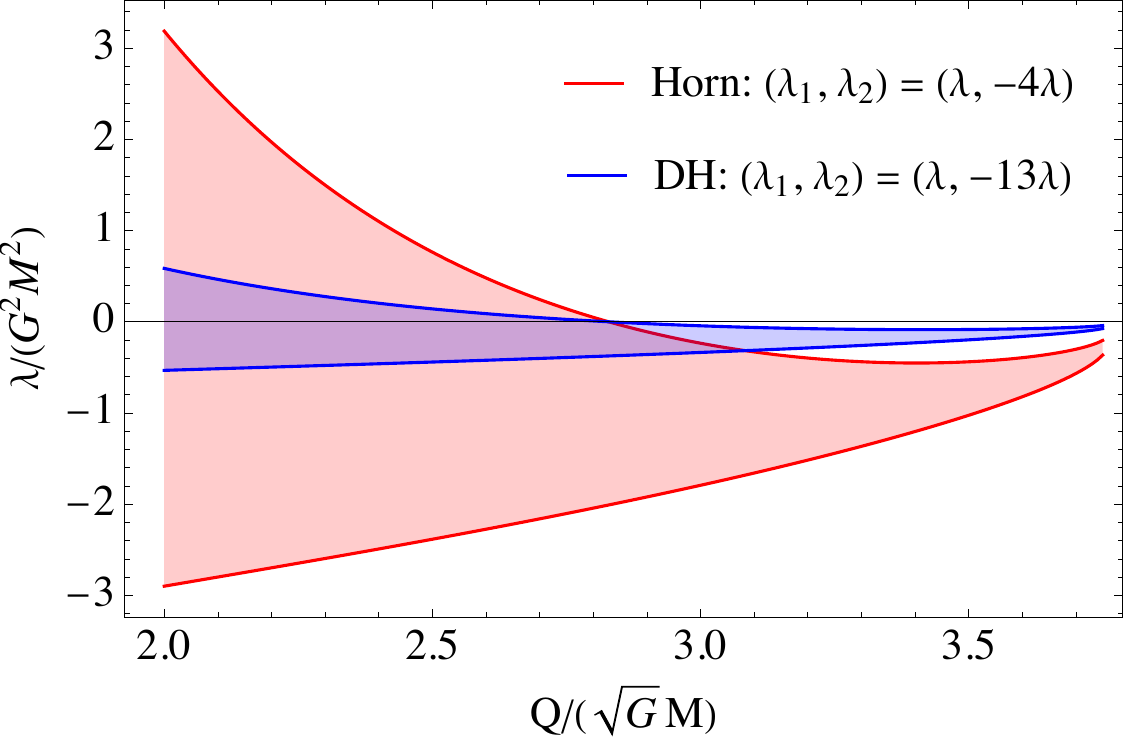}
\caption{\label{fig:bound} Allowed range of the $RF^2$ coupling constant as a function of the charge for the Horndeski (red) and DH (blue) combination from the black hole shadow observation of Sgr A* whose mass is estimated as $M = 4.0 \times 10^6M_\odot$ or $GM = 5.92 \times 10^6$km~\cite{EventHorizonTelescope:2022wkp}.
}
\end{figure}

Figure~\ref{fig:bound} shows bounds on the $RF^2$ coupling constant against $Q/(\sqrt{G}M)$ from black hole observation of Sgr A*. We present the allowed region in the parameter space for the Horndeski and DH combinations of the coupling constants \footnote{Although we ruled out the DH theory from black hole thermodynamics/weak gravity conjecture considerations, we are being agnostic about such constraints in this section and carry out order of magnitude estimates for observational bounds on the theory.}. A RN black hole ($\lambda = 0$) is consistent with the observation when $Q / (\sqrt{G}M) < 2.83$  while $\lambda$ needs to be non-vanishing (and negative) when $Q / (\sqrt{G}M)$ is above this threshold. We note that Eq.~\eqref{eq:M_ext} for the mass in the extremal configuration and the weak gravity conjecture suggests $\lambda < 0$ for the two combinations of parameters in the figure.
We also note that the bound is obtained within the small coupling approximation of $\lambda \ll G^2 M^2$, which means that the DH one should be valid while the Horndeski one may need to be understood as an order of magnitude estimate.

\section{Conclusions and Discussions}

In this paper we addressed the corrections to RN black holes due to higher-derivative $RF^2$ terms. Such terms arise naturally  in the context of effective field theories, either induced by quantum loops or generated by string theory $\alpha'$ corrections. We used QFT methods to obtain the leading order post-Minkowskian correction. We verified these results by solving the modified Einstein-Maxwell field equations in closed form, to all orders in Newton's constant $G$. We discussed the thermodynamic properties of the perturbed black holes. The first law holds. We derived the entropy of the perturbed black holes using both the first law and the Iyer-Wald formula, and found agreement. We obtained a constraint on the couplings of the $RF^2$ terms by requiring that the extremal RN solution remains regular in the presence of the higher derivative corrections. 
The same condition is obtained if  enforcing the  weak gravity conjecture. This constraint rules out the DH theory as a stand-alone extension of Einstein-Maxwell. In the case of Horndeski theory, the constraint fixes the sign of the overall coefficient $\gamma$ in \eqref{horndeski}, namely $\gamma<0$. We also found that the difference between the extremal higher-derivative black hole entropy  and the extremal RN black hole entropy is not necessarily positive when the weak gravity conjecture is satisfied, in contradiction to \cite{Cheung:2018cwt}.

We also presented bounds on the $RF^2$ coupling constants from black hole shadow observations of Sgr A*. In particular, we found bounds on the dimensionless coupling constant of $\lambda/(G^2M^2)$ to be less than $\mathcal{O}(1)$ when $2 < Q/(\sqrt{G}M)< 3.76$. If one converts these into dimensionful bounds on $\lambda$, they become of order $10^{13}$km$^2$, and hence are very weak bounds. For comparison, Shapiro time delay measurements by Cassini placed bounds for $\lambda_1$ (independent of $\lambda_{2,3}$) of order $10^{7}$km$^2$~\cite{HertzbergNathanSemaan2025}.
A similar analysis can be performed with the result from M87, though the bound on $\lambda$ will be even weaker as the mass is $10^3$ times larger than that of Sgr A*.

Stronger bounds should come from observations of black holes with smaller masses, such as gravitational wave observations of stellar-mass binary black hole coalesences. 
It would be interesting to derive how gravitational waveforms from a charged black hole binary inspiral are modified for $RF^2$ from RN. One can also derive corrections to the quasi-normal mode frequencies by carrying out a black hole perturbation analysis on the $RF^2$ black hole background found in this paper (a similar analysis was done in~\cite{delRio:2024kms} within the DH $RF^2$ theory). Another interesting avenue for future work is to derive the tidal Love number for the $RF^2$ black hole (the Love number for RN black holes has been shown to vanish~\cite{Rai:2024lho}) and compare with bounds from gravitational-wave observations (e.g.~\cite{Narikawa:2021pak}). One can follow Katagiri \emph{et al}.~\cite{Katagiri:2024fpn} that provides a formulation for computing black hole Love numbers in theories beyond General Relativity where the corrections are treated perturbatively.

There are a few other immediate directions for extending the present work. Since for simplicity we discussed here only the electrically charged black holes, it would be interesting to obtain the equivalent magnetically charged black holes, or the dyonic black holes. In this case, the challenge is to construct the magnetic monopole worldline QFT. Of note is that the Einstein-Maxwell equations are invariant under swapping the electric and magnetic fields (electric-magnetic duality), and so the Reissner-Nordstr\"om geometry remains the same regardless of whether it is sourced by a point-like electric or magnetic charge. Amusingly, the magnetically charged RN black holes were used to construct traversable wormholes in \cite{Maldacena:2018gjk}. On the other hand, the addition of higher-derivative terms of type $RF^2$ breaks this symmetry, and we expect that the perturbed magnetically charged RN black holes have a geometry different from that of the electrically charged RN black holes discussed here.

Another natural extension would be to include rotation (non-zero angular momentum) and find the higher curvature generalization of the Kerr--Newman metric. Of particular interest would be investigating whether the Newman--Janis algorithm holds in the presence of higher curvature terms; in~\cite{Arkani-Hamed:2019ymq}, the authors derived  the Kerr metric from the Schwarzschild metric by an exponentiated spin operator, providing a complementary field theoretic description of the Newman--Janis algorithm~\cite{Newman:1965tw}. An encouraging sign comes from \cite{Burger:2019wkq}, where
the work of \cite{Arkani-Hamed:2019ymq} was extended in the presence of generic cubic curvature $R^3$ perturbations.

\acknowledgments

S.A. and K.Y. acknowledge support from NSF Grant PHYS-2339969. K.Y. also acknowledges support from NSF Grant PHY-2309066.


\appendix
\section{Integrals}
\begin{equation}\label{eq:FT}
\int \frac{d^d \boldsymbol{q}}{(2 \pi)^d} \frac{e^{i \boldsymbol{q} \cdot \boldsymbol{x}}}{|\boldsymbol{q}|^n}
= \frac{1}{(4 \pi)^{d / 2}} \frac{\Gamma\left(\frac{d-n}{2}\right)}{\Gamma\left(\frac{n}{2}\right)}\left(\frac{2}{r}\right)^{d-n}
\end{equation}

\begin{equation}\label{eq:scalaroneloop}
J_{(1)}(\vb{q}^2) = \int \frac{d^d \boldsymbol{k}}{(2 \pi)^d} \frac{1}{\boldsymbol{k}^2(\boldsymbol{k}+\boldsymbol{q})^2}
= \frac{\Gamma\left(2-\frac{d}{2}\right) \qty(\Gamma\qty(\frac{d-2}{2}))^2}{(4 \pi)^{\frac{d}{2}} \Gamma(d-2)}|\boldsymbol{q}|^{d}
\end{equation}

\begin{equation}\label{eq:tensoroneloop}
\int \frac{d^d \boldsymbol{q}}{(2 \pi)^d} e^{i \boldsymbol{q} \cdot \boldsymbol{x}} \frac{\boldsymbol{q}_i \boldsymbol{q}_j}{|\boldsymbol{q}|^{n+2}}
= \left(\int \frac{d^d \boldsymbol{q}}{(2 \pi)^d} \frac{e^{i \boldsymbol{q} \cdot \boldsymbol{x}}}{|\boldsymbol{q}|^n}\right)
\left(\frac{1}{n} \delta_{i j}+\frac{n-d}{n} \,\frac{\boldsymbol{x}_i \boldsymbol{x}_j}{r^2}\right)
\end{equation}

\section{Iyer--Wald 
Entropy}\label{AppendixIW}
We consider a static, stationary metric and vector field written in the form given by Eqs.~\eqref{eq:metric_ans} --~\eqref{eq:metric_f}. The outer and inner event horizons are given in Eqs.~\eqref{eq:rpm} --~\eqref{eq:r_inner}, which are solved by the condition $f(r_{\text{H}})=0$. For this analysis, we split the manifold into $\mathcal M=\mathcal M_2\times S_2$ where $S_2$ is the two-sphere spanned by $\{\theta,\phi\}$ and $\mathcal M_2$ is the submanifold spanned by coordinate indices $\{t,r\}$. Let us first define the Iyer--Wald entropy as well as the geometric structures needed for the analysis before proceeding to solve for the entropy of the black hole solution.

The Iyer--Wald entropy~\cite{Iyer:1994ys} is
\begin{equation}
    S_{\rm IW} = -2\pi \int_{\mathcal H} d^2x\,\sqrt{\gamma}\, E_R^{\mu\nu\rho\sigma} \varepsilon_{\mu\nu}\varepsilon_{\rho\sigma},
    \label{eq:IW_entropy}
\end{equation}
where we define the tensor conjugate to the curvature tensor,
\begin{equation}
    E_R^{\mu\nu\rho\sigma} := \frac{\partial \mathcal L}{\partial R_{\mu\nu\rho\sigma}} . \label{eq:ER_def}
\end{equation}
Above, $\mathcal H$ is the horizon on which we integrate over the remaining $S_2$ degree of freedom, $\gamma$ is the metric determinant of the $S_2$ metric, and $\epsilon_{\mu\nu}$ is the volume element corresponding to $\mathcal M_2$. In this calculation, we take the outer horizon to be $\mathcal H$. This means that for the remainder of the entropy calculation, the radial coordinate will be evaluated at the outer horizon $r_{\text{outer\,H}}$. For the metric \eqref{eq:metric_ans}, we have
\begin{equation}
    \sqrt{\gamma}=r_{\text{outer\,H}}^2\sin\theta, \qquad A_{\mathcal H}=4\pi r_{\text{outer\,H}}^2,
\end{equation}
and the normalized binormal tensor given by
\begin{equation}
    \varepsilon_{\mu\nu}\varepsilon^{\mu\nu}=-2, \qquad \varepsilon_{tr}=N(r_{\text{outer\,H}}), \qquad \varepsilon_{rt}=-N(r_{\text{outer\,H}}). \label{eq:binormal_Nf}
\end{equation}
Therefore,
\begin{equation}
    E_R^{\mu\nu\rho\sigma} \varepsilon_{\mu\nu}\varepsilon_{\rho\sigma} = 4N(r_{\text{outer\,H}})^2 E_R^{trtr}(r_{\text{outer\,H}}),
    \label{eq:ER_binormal_contraction}
\end{equation}
and hence
\begin{equation}
    S_{\rm IW} = -8\pi A_{\mathcal H} N(r_{\text{outer\,H}})^2 E_R^{trtr}(r_{\text{outer\,H}}).
    \label{eq:entropy_Etrtr}
\end{equation}
Only the terms with explicit curvature dependence contribute directly to
$E_R^{\mu\nu\rho\sigma}$. Thus
\begin{align}
    E_R^{\mu\nu\rho\sigma}
    &= \frac{1}{32\pi G} \left( g^{\mu\rho}g^{\nu\sigma} - g^{\mu\sigma}g^{\nu\rho} \right)  +\epsilon\bigg[\lambda_1 F^{\mu\nu}F^{\rho\sigma} +\lambda_3 \left(F_{\alpha\beta}F^{\alpha\beta}\right) g^{\mu[\rho}g^{\sigma]\nu} \nonumber\\
    &\quad +\frac{\lambda_2}{4} \left( g^{\mu\rho}F^\nu{}_\alpha F^{\sigma\alpha} -g^{\mu\sigma}F^\nu{}_\alpha F^{\rho\alpha} -g^{\nu\rho}F^\mu{}_\alpha F^{\sigma\alpha} +g^{\nu\sigma}F^\mu{}_\alpha F^{\rho\alpha} \right)\bigg],
    \label{eq:ER_perturbative}
\end{align}
where the $\epsilon$ was introduced as an order-counting parameter. We will be consistently working to linear order in the $\lambda_i$, that is, we will truncate to linear order in $\epsilon$.

For the purely electric ansatz $\mathbf A=\Phi (r)\mathbf dt$, we have
\begin{equation}
    F_{tr}=-\Phi '(r), \quad F^{tr}=\frac{\Phi '(r)}{N(r)^2},
    \label{eq:F_components}
\end{equation}
and can find the useful compositions
\begin{equation}
F_{\mu\nu}F^{\mu\nu} = -\frac{2\Phi '(r)^2}{N(r)^2},\quad F^{t\alpha}F^t{}_{\alpha}=\frac{1}{f(r)}\left(\frac{\Phi'(r)}{N(r)^2}\right)^2,\quad F^{r\alpha}F^r{}_{\alpha}=-f(r)N(r)^2\left(\frac{\Phi'(r)}{N(r)^2}\right)^2.
\end{equation}
Using these along with metric components,
\begin{equation}
    g^{tt}=-\frac{1}{N(r)^2 f(r)}=\frac{1}{g_{tt}},
    \qquad
    g^{rr}=f(r)=\frac{1}{g_{rr}},
    \qquad
    g^{tt}g^{rr}=-\frac{1}{N(r)^2},
\end{equation}
one obtains the explicit curvature-conjugate tensor component
\begin{align}
    E_R^{trtr} &= -\frac{1}{32\pi G\,N(r)^2} +\epsilon\bigg[\lambda_1\frac{\Phi '(r)^2}{N(r)^4} +\frac{\lambda_2}{2}\frac{\Phi '(r)^2}{N(r)^4} +\lambda_3\frac{\Phi '(r)^2}{N(r)^4}\bigg]
    \nonumber\\
    &=
    -\frac{1}{32\pi G\,N(r)^2} + \epsilon\left( \lambda_1+\frac{\lambda_2}{2}+\lambda_3 \right) \frac{\Phi '(r)^2}{N(r)^4}.
    \label{eq:Etrtr_explicit}
\end{align}
Therefore the entropy is
\begin{align}
    S_{\rm IW} &= -8\pi A_{\mathcal H}N(r_{\text{outer\,H}})^2 \left[ -\frac{1}{32\pi G\,N(r_{\text{outer\,H}})^2} + \left( \lambda_1+\frac{\lambda_2}{2}+\lambda_3 \right) \frac{\Phi '(r_{\text{outer\,H}})^2}{N(r_{\text{outer\,H}})^4} \right]
    \nonumber\\
    &= \frac{A_{\mathcal H}}{4G} - 8\pi A_{\mathcal H} \left( \lambda_1+\frac{\lambda_2}{2}+\lambda_3 \right) \frac{\Phi '(r_{\text{outer\,H}})^2}{N(r_{\text{outer\,H}})^2}.
    \label{eq:entropy_Nf_final}
\end{align}
Since $A_{\mathcal H}=4\pi r_{\text{outer\,H}}^2$, this becomes
\begin{equation}
    S_{\rm IW} = \frac{\pi r_{\text{outer\,H}}^2}{G} - 32\pi^2 r_{\text{outer\,H}}^2 \,\epsilon\left( \lambda_1+\frac{\lambda_2}{2}+\lambda_3 \right) \frac{\Phi '(r_{\text{outer\,H}})^2}{N(r_{\text{outer\,H}})^2}.
    \label{eq:entropy_Nf_final_area}
\end{equation}

At linear order in the higher-derivative expansion, this can be written as
\begin{align}
    S_{\rm IW}
    &=
    \frac{\pi r_+^2}{G} + \epsilon\,\frac{2\pi r_+}{G}\,\delta r_{\text{outer\,H}} - 32\pi^2 r_+^2 \epsilon \left( \lambda_1+\frac{\lambda_2}{2}+\lambda_3 \right) \left[
    \Phi_0'(r_+) \right]^2 +\mathcal O(\epsilon^2),
    \label{eq:entropy_linear_hd}
\end{align}
where
\begin{equation}
    \Phi_0(r)=\frac{Q}{4\pi r},
    \qquad
    \Phi_0'(r)=-\frac{Q}{4\pi r^2}.
\end{equation}
Thus
\begin{equation}
    S_{\rm IW} = \frac{\pi r_+^2}{G} + \epsilon\,\frac{2\pi r_+}{G}\,\delta r_{\text{outer\,H}} - 2 \epsilon\left( \lambda_1+\frac{\lambda_2}{2}+\lambda_3 \right) \frac{Q^2}{r_+^2} +\mathcal O(\epsilon^2).
    \label{eq:entropy_linear_explicit}
\end{equation}
Writing this in terms of the horizon area, truncating to leading order in $\epsilon$ (and subsequently removing the perturbative order counting parameter $\epsilon$), we obtain the following Iyer-Wald entropy
\begin{equation}
    S_{IW}=\frac{A_{\mathcal H}}{4 G}-8\pi\left(\lambda_1+\frac{\lambda_2}{2}+\lambda_3\right)\frac{Q^2}{A_{\mathcal H}}.
\end{equation}
The Iyer-Wald entropy matches exactly with Eq.~\eqref{entropy} which was derived using the first law.


\bibliography{References}

@article{Newman:1965tw,
    author = "Newman, E. T. and Janis, A. I.",
    title = "{Note on the Kerr spinning particle metric}",
    doi = "10.1063/1.1704350",
    journal = "J. Math. Phys.",
    volume = "6",
    pages = "915--917",
    year = "1965"
}

@article{Arkani-Hamed:2019ymq,
    author = "Arkani-Hamed, Nima and Huang, Yu-tin and O'Connell, Donal",
    title = "{Kerr black holes as elementary particles}",
    eprint = "1906.10100",
    archivePrefix = "arXiv",
    primaryClass = "hep-th",
    reportNumber = "NCTS-TH/1905",
    doi = "10.1007/JHEP01(2020)046",
    journal = "JHEP",
    volume = "01",
    pages = "046",
    year = "2020"
}

@article{Donoghue1994a,
  author       = {Donoghue, John F.},
  title        = "{Leading quantum correction to the Newtonian potential}",
  journal      = {Physical Review Letters},
  volume       = {72},
  pages        = {2996--2999},
  year         = {1994},
  eprint       = {gr-qc/9310024},
  archivePrefix = {arXiv},
  doi          = {10.1103/PhysRevLett.72.2996}
}

@article{BjerrumBohr2003,
  author       = {Bjerrum-Bohr, N. E. J. and Donoghue, John F. and Holstein, Barry R.},
  title        = "{Quantum corrections to the Schwarzschild and Kerr metrics}",
  journal      = {Physical Review D},
  volume       = {68},
  pages        = {084005},
  year         = {2003},
  eprint       = {hep-th/0211071},
  archivePrefix = {arXiv},
  doi          = {10.1103/PhysRevD.68.084005}
}

@article{DonoghueGarbrecht2002,
  author = {Donoghue, J. F. and Holstein, B. R. and Garbrecht, B. and Konstandin, T.},
  title = "{Quantum Corrections to the Reissner-Nordstrom and Kerr-Newman Metrics}",
  journal = {Phys. Lett. B},
  volume = {529},
  pages = {132--142},
  year = {2002},
  doi = {10.1016/S0370-2693(02)01246-7},
  eprint = {hep-th/0112237}
}

@article{Horndeski1976,
  author = {Horndeski, G. W.},
  title = "{Conservation of Charge and the Einstein-Maxwell Field Equations}",
  journal = {J. Math. Phys.},
  volume = {17},
  pages = {1980},
  year = {1976}
}

@article{DrummondHathrell1980,
  author = {Drummond, I. T. and Hathrell, S. J.},
  title = "{QED vacuum polarization in a background gravitational field and its effect on the velocity of photons}",
  journal = {Phys. Rev. D},
  volume = {22},
  pages = {343},
  year = {1980}
}

@article{KatsMotlPadi2007,
  author = {Kats, Y. and Motl, L. and Padi, M.},
  title = {Higher-order corrections to mass-charge relation of extremal black holes},
  journal = {JHEP},
  volume = {12},
  pages = {068},
  year = {2007},
  eprint = {hep-th/0606100},
  archivePrefix = {arXiv}
}

@article{HertzbergNathanSemaan2025,
  author = {Hertzberg, M. P. and Nathan, R. and Semaan, S. E.},
  title = "{Solar System constraints on light propagation from higher-derivative corrections to general relativity and implications for fundamental physics}",
  journal = {Phys. Rev. D},
  volume = {112},
  pages = {064078},
  year = {2025},
  eprint = {2503.19236},
  archivePrefix = {arXiv}
}

@article{Bastianelli2009,
  author = {Bastianelli, F. and Davila, J. M. and Schubert, C.},
  title = "{Gravitational corrections to the Euler--Heisenberg Lagrangian}",
  journal = {JHEP},
  volume = {03},
  pages = {086},
  year = {2009},
  eprint = {0812.4849},
  archivePrefix = {arXiv}
}

@article{Cremonini:2009ih,
    author = "Cremonini, Sera and Liu, James T. and Szepietowski, Phillip",
    title = "{Higher Derivative Corrections to R-charged Black Holes: Boundary Counterterms and the Mass-Charge Relation}",
    eprint = "0910.5159",
    archivePrefix = "arXiv",
    primaryClass = "hep-th",
    reportNumber = "MCTP-09-51, MIFP-09-44, DAMTP-2009-72",
    doi = "10.1007/JHEP03(2010)042",
    journal = "JHEP",
    volume = "03",
    pages = "042",
    year = "2010"
}

@article{DOnofrio:2022cvn,
    author = "D'Onofrio, Simone and Fragomeno, Federica and Gambino, Claudio and Riccioni, Fabio",
    title = {{The Reissner-Nordstr{\"o}m-Tangherlini solution from scattering amplitudes of charged scalars}},
    eprint = "2207.05841",
    archivePrefix = "arXiv",
    primaryClass = "hep-th",
    doi = "10.1007/JHEP09(2022)013",
    journal = "JHEP",
    volume = "09",
    pages = "013",
    year = "2022"
}

@article{Strassler:1992zr,
    author = "Strassler, Matthew J.",
    title = "{Field theory without Feynman diagrams: One loop effective actions}",
    eprint = "hep-ph/9205205",
    archivePrefix = "arXiv",
    reportNumber = "SLAC-PUB-5757",
    doi = "10.1016/0550-3213(92)90098-V",
    journal = "Nucl. Phys. B",
    volume = "385",
    pages = "145--184",
    year = "1992"
}

@article{Ajith:2024fna,
    author = "Ajith, Siddarth and Du, Yuchen and Rajagopal, Ravisankar and Vaman, Diana",
    title = "{Worldline formalism, eikonal expansion and the classical limit of scattering amplitudes}",
    eprint = "2409.17866",
    archivePrefix = "arXiv",
    primaryClass = "hep-th",
    doi = "10.1016/j.nuclphysb.2026.117367",
    journal = "Nucl. Phys. B",
    volume = "1025",
    pages = "117367",
    year = "2026"
}

@article{Mogull:2020sak,
    author = "Mogull, Gustav and Plefka, Jan and Steinhoff, Jan",
    title = "{Classical black hole scattering from a worldline quantum field theory}",
    eprint = "2010.02865",
    archivePrefix = "arXiv",
    primaryClass = "hep-th",
    reportNumber = "UUITP-37/20, HU-EP-20/22-RTG",
    doi = "10.1007/JHEP02(2021)048",
    journal = "JHEP",
    volume = "02",
    pages = "048",
    year = "2021"
}

@article{delRio:2024kms,
    author = "del R{\'\i}o, Adri{\'a}n and Ester, Evelyn-Andreea",
    title = "{Electrically charged black hole solutions in semiclassical gravity and dynamics of linear perturbations}",
    eprint = "2401.08783",
    archivePrefix = "arXiv",
    primaryClass = "gr-qc",
    doi = "10.1103/PhysRevD.109.105022",
    journal = "Phys. Rev. D",
    volume = "109",
    number = "10",
    pages = "105022",
    year = "2024"
}

@article{Du:2024rkf,
    author = "Du, Yuchen and Ajith, Siddarth and Rajagopal, Ravisankar and Vaman, Diana",
    title = "{Worldline proof of eikonal exponentiation}",
    eprint = "2409.12895",
    archivePrefix = "arXiv",
    primaryClass = "hep-th",
    doi = "10.1007/JHEP09(2025)161",
    journal = "JHEP",
    volume = "09",
    pages = "161",
    year = "2025"
}

@article{Campanelli:1994sj,
    author = "Campanelli, Manuela and Lousto, C. O. and Audretsch, J.",
    title = "{A Perturbative method to solve fourth order gravity field equations}",
    eprint = "gr-qc/9401013",
    archivePrefix = "arXiv",
    reportNumber = "UAB-FT-327",
    doi = "10.1103/PhysRevD.49.5188",
    journal = "Phys. Rev. D",
    volume = "49",
    pages = "5188--5193",
    year = "1994"
}

@article{Campanelli:1994jt,
    author = "Campanelli, Manuela and Lousto, C. O. and Audretsch, J.",
    title = "{Perturbative metric of charged black holes in quadratic gravity}",
    eprint = "gr-qc/9412001",
    archivePrefix = "arXiv",
    reportNumber = "UAB-FT-354",
    doi = "10.1103/PhysRevD.51.6810",
    journal = "Phys. Rev. D",
    volume = "51",
    pages = "6810--6815",
    year = "1995"
}

@article{Mougiakakos:2024nku,
    author = "Mougiakakos, Stavros and Vanhove, Pierre",
    title = "{Schwarzschild Metric from Scattering Amplitudes to All Orders in GN}",
    eprint = "2405.14421",
    archivePrefix = "arXiv",
    primaryClass = "hep-th",
    reportNumber = "IPhT-t24/010",
    doi = "10.1103/PhysRevLett.133.111601",
    journal = "Phys. Rev. Lett.",
    volume = "133",
    number = "11",
    pages = "111601",
    year = "2024"
}

@article{Arkani-Hamed:2006emk,
    author = "Arkani-Hamed, Nima and Motl, Lubos and Nicolis, Alberto and Vafa, Cumrun",
    title = "{The String landscape, black holes and gravity as the weakest force}",
    eprint = "hep-th/0601001",
    archivePrefix = "arXiv",
    reportNumber = "HUTP-05-A0057",
    doi = "10.1088/1126-6708/2007/06/060",
    journal = "JHEP",
    volume = "06",
    pages = "060",
    year = "2007"
}

@article{EventHorizonTelescope:2022wkp,
    author = "Akiyama, Kazunori and others",
    collaboration = "Event Horizon Telescope",
    title = "{First Sagittarius A* Event Horizon Telescope Results. I. The Shadow of the Supermassive Black Hole in the Center of the Milky Way}",
    eprint = "2311.08680",
    archivePrefix = "arXiv",
    primaryClass = "astro-ph.HE",
    doi = "10.3847/2041-8213/ac6674",
    journal = "Astrophys. J. Lett.",
    volume = "930",
    number = "2",
    pages = "L12",
    year = "2022"
}

@article{EventHorizonTelescope:2019dse,
    author = "Akiyama, Kazunori and others",
    collaboration = "Event Horizon Telescope",
    title = "{First M87 Event Horizon Telescope Results. I. The Shadow of the Supermassive Black Hole}",
    eprint = "1906.11238",
    archivePrefix = "arXiv",
    primaryClass = "astro-ph.GA",
    doi = "10.3847/2041-8213/ab0ec7",
    journal = "Astrophys. J. Lett.",
    volume = "875",
    pages = "L1",
    year = "2019"
}

@article{Perlick:2021aok,
    author = "Perlick, Volker and Tsupko, Oleg Yu.",
    title = "{Calculating black hole shadows: Review of analytical studies}",
    eprint = "2105.07101",
    archivePrefix = "arXiv",
    primaryClass = "gr-qc",
    doi = "10.1016/j.physrep.2021.10.004",
    journal = "Phys. Rept.",
    volume = "947",
    pages = "1--39",
    year = "2022"
}

@article{Vagnozzi:2022moj,
    author = "Vagnozzi, Sunny and others",
    title = "{Horizon-scale tests of gravity theories and fundamental physics from the Event Horizon Telescope image of Sagittarius A}",
    eprint = "2205.07787",
    archivePrefix = "arXiv",
    primaryClass = "gr-qc",
    reportNumber = "UCI-HEP-TR-2022-07",
    doi = "10.1088/1361-6382/acd97b",
    journal = "Class. Quant. Grav.",
    volume = "40",
    number = "16",
    pages = "165007",
    year = "2023"
}

@article{Katagiri:2024fpn,
    author = "Katagiri, Takuya and Cardoso, Vitor and Ikeda, Tact and Yagi, Kent",
    title = "{Tidal response beyond vacuum general relativity with a canonical definition}",
    eprint = "2410.02531",
    archivePrefix = "arXiv",
    primaryClass = "gr-qc",
    reportNumber = "RUP-24-19",
    doi = "10.1103/PhysRevD.111.084081",
    journal = "Phys. Rev. D",
    volume = "111",
    number = "8",
    pages = "084081",
    year = "2025"
}

@article{Rai:2024lho,
    author = "Rai, Mudit and Santoni, Luca",
    title = {{Ladder symmetries and Love numbers of Reissner-Nordstr{\"o}m black holes}},
    eprint = "2404.06544",
    archivePrefix = "arXiv",
    primaryClass = "gr-qc",
    doi = "10.1007/JHEP07(2024)098",
    journal = "JHEP",
    volume = "07",
    pages = "098",
    year = "2024"
}

@article{Narikawa:2021pak,
    author = "Narikawa, Tatsuya and Uchikata, Nami and Tanaka, Takahiro",
    title = "{Gravitational-wave constraints on the GWTC-2 events by measuring the tidal deformability and the spin-induced quadrupole moment}",
    eprint = "2106.09193",
    archivePrefix = "arXiv",
    primaryClass = "gr-qc",
    doi = "10.1103/PhysRevD.104.084056",
    journal = "Phys. Rev. D",
    volume = "104",
    number = "8",
    pages = "084056",
    year = "2021",
    note = "[Erratum: Phys.Rev.D 111, 089903 (2025)]"
}

@article{Iyer:1994ys,
    author = "Iyer, Vivek and Wald, Robert M.",
    title = "{Some properties of Noether charge and a proposal for dynamical black hole entropy}",
    eprint = "gr-qc/9403028",
    archivePrefix = "arXiv",
    doi = "10.1103/PhysRevD.50.846",
    journal = "Phys. Rev. D",
    volume = "50",
    pages = "846--864",
    year = "1994"
}

@article{Burger:2019wkq,
    author = "Burger, Daniel J. and Emond, William T. and Moynihan, Nathan",
    title = "{Rotating Black Holes in Cubic Gravity}",
    eprint = "1910.11618",
    archivePrefix = "arXiv",
    primaryClass = "hep-th",
    doi = "10.1103/PhysRevD.101.084009",
    journal = "Phys. Rev. D",
    volume = "101",
    number = "8",
    pages = "084009",
    year = "2020"
}

@article{Maldacena:2018gjk,
    author = "Maldacena, Juan and Milekhin, Alexey and Popov, Fedor",
    title = "{Traversable wormholes in four dimensions}",
    eprint = "1807.04726",
    archivePrefix = "arXiv",
    primaryClass = "hep-th",
    doi = "10.1088/1361-6382/acde30",
    journal = "Class. Quant. Grav.",
    volume = "40",
    number = "15",
    pages = "155016",
    year = "2023"
}

@article{Cheung:2018cwt,
    author = "Cheung, Clifford and Liu, Junyu and Remmen, Grant N.",
    title = "{Proof of the Weak Gravity Conjecture from Black Hole Entropy}",
    eprint = "1801.08546",
    archivePrefix = "arXiv",
    primaryClass = "hep-th",
    reportNumber = "CALT-TH-2018-007",
    doi = "10.1007/JHEP10(2018)004",
    journal = "JHEP",
    volume = "10",
    pages = "004",
    year = "2018"
}

@article{Heisenberg:1936nmg,
    author = "Heisenberg, W. and Euler, H.",
    title = "{Consequences of Dirac's theory of positrons}",
    eprint = "physics/0605038",
    archivePrefix = "arXiv",
    doi = "10.1007/BF01343663",
    journal = "Z. Phys.",
    volume = "98",
    number = "11-12",
    pages = "714--732",
    year = "1936"
}

@article{Reissner1916,
  author  = {Reissner, Hans},
  title   = "{{\"U}ber die Eigengravitation des elektrischen Feldes nach der Einsteinschen Theorie}",
  journal = {Annalen der Physik},
  volume  = {355},
  number  = {9},
  pages   = {106--120},
  year    = {1916},
  doi     = {10.1002/andp.19163550905},
  bibcode = {1916AnP...355..106R}
}

@article{Nordstrom1918,
  author  = {Nordstr{\"o}m, Gunnar},
  title   = "{On the Energy of the Gravitation Field in Einstein's Theory}",
  journal = {Koninklijke Nederlandsche Akademie van Wetenschappen Proceedings},
  volume  = {20},
  number  = {2},
  pages   = {1238--1245},
  year    = {1918},
  bibcode = {1918KNAB...20.1238N}
}

@article{BardeenCarterHawking1973,
  author  = {Bardeen, J. M. and Carter, B. and Hawking, S. W.},
  title   = "{The Four Laws of Black Hole Mechanics}",
  journal = {Communications in Mathematical Physics},
  volume  = {31},
  number  = {2},
  pages   = {161--170},
  year    = {1973},
  doi     = {10.1007/BF01645742},
  bibcode = {1973CMaPh..31..161B}
}

@article{Bekenstein1973,
  author  = {Bekenstein, Jacob D.},
  title   = "{Black Holes and Entropy}",
  journal = {Physical Review D},
  volume  = {7},
  number  = {8},
  pages   = {2333--2346},
  year    = {1973},
  doi     = {10.1103/PhysRevD.7.2333}
}

@article{Hawking1975,
  author  = {Hawking, S. W.},
  title   = "{Particle Creation by Black Holes}",
  journal = {Communications in Mathematical Physics},
  volume  = {43},
  number  = {3},
  pages   = {199--220},
  year    = {1975},
  doi     = {10.1007/BF02345020}
}

@article{Schubert_2001,
   title={Perturbative quantum field theory in the string-inspired formalism},
   volume={355},
   ISSN={0370-1573},
   url={http://dx.doi.org/10.1016/S0370-1573(01)00013-8},
   DOI={10.1016/s0370-1573(01)00013-8},
   number={2-3},
   journal={Physics Reports},
   publisher={Elsevier BV},
   author={Schubert, Christian},
   year={2001},
   month=Dec, pages={73–234} }

@misc{edwards2019quantummechanicalpathintegrals,
      title={Quantum mechanical path integrals in the first quantised approach to quantum field theory}, 
      author={James P. Edwards and Christian Schubert},
      year={2019},
      eprint={1912.10004},
      archivePrefix={arXiv},
      primaryClass={hep-th},
      url={https://arxiv.org/abs/1912.10004}, 
}

@article{Mougiakakos_2021,
   title="{Schwarzschild-Tangherlini metric from scattering amplitudes in various dimensions}",
   volume={103},
   ISSN={2470-0029},
   url={http://dx.doi.org/10.1103/PhysRevD.103.026001},
   DOI={10.1103/physrevd.103.026001},
   number={2},
   journal={Physical Review D},
   publisher={American Physical Society (APS)},
   author={Mougiakakos, Stavros and Vanhove, Pierre},
   year={2021},
   month=Jan }

@article{Shore:2003zc,
    author = "Shore, Graham M.",
    title = "{Quantum gravitational optics}",
    eprint = "gr-qc/0304059",
    archivePrefix = "arXiv",
    reportNumber = "SWAT-03-375",
    doi = "10.1080/00107510310001617106",
    journal = "Contemp. Phys.",
    volume = "44",
    pages = "503--521",
    year = "2003"
}

@article{McPeak:2021tvu,
    author = "McPeak, Brian",
    title = "{Higher-derivative corrections to black hole entropy at zero temperature}",
    eprint = "2112.13433",
    archivePrefix = "arXiv",
    primaryClass = "hep-th",
    doi = "10.1103/PhysRevD.105.L081901",
    journal = "Phys. Rev. D",
    volume = "105",
    number = "8",
    pages = "L081901",
    year = "2022"
}

@article{Hamada:2018dde,
    author = "Hamada, Yuta and Noumi, Toshifumi and Shiu, Gary",
    title = "{Weak Gravity Conjecture from Unitarity and Causality}",
    eprint = "1810.03637",
    archivePrefix = "arXiv",
    primaryClass = "hep-th",
    reportNumber = "CCTP-2018-12, ITCP-IPP 2018/9, KOBE-COSMO-18-08, MAD-TH-18-05",
    doi = "10.1103/PhysRevLett.123.051601",
    journal = "Phys. Rev. Lett.",
    volume = "123",
    number = "5",
    pages = "051601",
    year = "2019"
}

@article{Alonzo-Artiles:2026wbe,
    author = "Alonzo-Artiles, Allan and Kraus, Manfred",
    title = {{Classical Hamiltonian of Reissner-Nordstr{\"o}m black holes at second post-Minkowskian order from scattering amplitudes}},
    eprint = "2603.15933",
    archivePrefix = "arXiv",
    primaryClass = "hep-th",
    doi = "10.1103/zpwl-r7nm",
    journal = "Phys. Rev. D",
    volume = "113",
    number = "12",
    pages = "124061",
    year = "2026"
}
\end{document}